\documentclass[reprint,floatfix,sor,footinbib,unsortedaddress]{revtex4-1}

\bibpunct[, ]{[}{]}{,}{n}{,}{,}

\renewcommand\bibnumfmt[1]{[#1] }

\usepackage{graphics}
\usepackage{graphicx}
\usepackage{psfrag}
\usepackage{latexsym}
\usepackage{textcomp}
\usepackage{amssymb}
\usepackage{amsmath}
\usepackage{hyperref} 
\usepackage{bm}
\usepackage[Euler]{upgreek}
\usepackage{longtable}
\usepackage{subfigure}

\usepackage{epsfig}
\usepackage{dcolumn}

\usepackage{multirow}
\usepackage{textcomp}

\DeclareMathAlphabet{\mathsf}{OT1}{phv}{b}{n}

\newcommand{\Vector}[1]{\ensuremath{\mathbf{#1}}}
\newcommand{\Tensor}[1]{\ensuremath{\mathsf{#1}}}

\newcommand{\bcal}[1]{\bm{\mathcal{#1}}}

\newcommand{\avg}[1]{\left< #1 \right>}

\newcommand{\Exp}[1]{{\rm e}^{#1}}

\newcommand{\crossVorg}{\ensuremath{%
         \setbox0=\hbox{$V$}
        V \kern-\wd0{\raise.3ex\hbox{$\relbar$}}}}

\newcommand{\crossVxx}[2]{%
	{\setbox0=\hbox{$#1#2V$}
         \setbox1=\hbox{$#1#2$}
         \setbox2=\hbox{$#1V$}
         \dimen1=\wd0
	 \advance\dimen1-\wd1
         \raise.2\ht0\hbox{$#1#2$}\kern-.4\wd0}}

\newcommand{\etal}{\textit{et~al.}\@}
\newcommand{\degree}{\ensuremath{^\circ}}
\newcommand{\degC}{\ensuremath{\degree\mathrm{C}}}

\usepackage{pifont}

\makeatletter 
\gdef\@ptsize{0}
\let\@currsize\normalsize 
\makeatother 
\usepackage{setspace} 

\begin{document}

\title{Parameter-free prediction of DNA dynamics in planar extensional flow of semidilute solutions}
\date{\today}
\author{Chandi Sasmal}
\affiliation{Department of Chemical Engineering, Monash University, Melbourne, VIC 3800, Australia}
\author{Kai-Wen Hsiao}
\affiliation{Department of Chemical \& Biomolecular Engineering, University of Illinois at Urbana-Champaign, Urbana, Illinois 61801, United States}
\author{Charles M. Schroeder}
\affiliation{Department of Chemical \& Biomolecular Engineering, University of Illinois at Urbana-Champaign, Urbana, Illinois 61801, United States}
\author{J. Ravi Prakash}
\email[Corresponding author: ]{ravi.jagadeeshan@monash.edu}
\affiliation{Department of Chemical Engineering, Monash University, Melbourne, VIC 3800, Australia}


\begin{abstract}
The dynamics of individual DNA molecules in semidilute solutions undergoing planar extensional flow is simulated using a multi-particle Brownian dynamics algorithm, which incorporates hydrodynamic and excluded volume interactions in the context of a coarse-grained bead-spring chain model for DNA. The successive fine-graining protocol~\cite{Sunthar,PrabhakarSFG}, in which simulation data acquired for bead-spring chains with increasing values of the number of beads $N_b$, is extrapolated to the number of Kuhn steps $N_\text{K}$ in DNA (while keeping key physical parameters invariant), is used to obtain parameter-free predictions for a range of Weissenberg numbers and Hencky strain units. A systematic comparison of simulation predictions is carried out with the experimental observations of~\citet{Kaiwen}, who have recently used single molecule techniques to investigate the dynamics of dilute and semidilute solutions of $\lambda$-phage DNA in planar extensional flow. In particular, they examine the response of individual chains to step-strain deformation followed by cessation of flow, thereby capturing both chain stretch and relaxation in a single experiment. The successive fine-graining technique is shown to lead to quantitatively accurate predictions of the experimental observations in the stretching and relaxation phases. Additionally, the transient chain stretch following a step strain deformation is shown to be much smaller in semidilute solutions than in dilute solutions, in agreement with experimental observations. 
\end{abstract}


\maketitle


\section{\label{sec:intro}Introduction}

Several studies of single molecules of fluorescently labelled DNA have been carried out in order to gain insight into the conformational evolution of polymer chains when subjected to a variety of flow fields~\cite{Amanda&Schroeder, Fluorescence, Danielle&Schroeder, Perkins, Smith&Chu, Leduc, SmithShear, HazenMixed, Wirtz, PerkinsRelax, Quake, CharlesCoilStretch, PerkinCon, SmithDif, Huber, SmithDifSca, PerkinsTetheredPoly, BabcockPRL, Teixeira, Harasim}. These studies have not only enabled the direct visual observation of `molecular individualism'~\cite{deGennesMol,Smith&Chu}, but have also proved to be of vital importance for the validation of molecular theories of polymer dynamics~\cite{Shaqfeh,Sunthar,LarsonExt,Jendrejack,Hsieh,CharlesExt,PerkinCon,SmithDif,Hur,SmithDifSca,Robertson}. Nearly all these investigations have been carried out in either the dilute or concentrated solution regimes, with only a few in the semidilute regime~\cite{Hur,BabcockPRL,Harasim,Huber,Steinberg}. Given the importance of semidilute polymer solutions, both from a fundamental and a practical~\cite{Nanofiber,Inkjet,Kozer} point of view, it is essential to gain an understanding of the fundamental physics that govern the dynamics of polymer molecules in this regime. In the dilute regime, single molecules studies have revealed the importance of properly accounting for hydrodynamic and excluded volume interactions in molecular theories~\cite{Shaqfeh,Sunthar,LarsonExt,Jendrejack,Hsieh,CharlesExt}. In semidilute solutions, however, it is known that these interactions gradually get screened with increasing monomer concentration~\cite{deGennes,Rubinstein,JainPRL}. The recent single molecule experiments of Hsiao \etal~on planar extensional flow of unentangled semidilute solutions of $\lambda$-phage DNA (reported in a companion paper~\cite{Kaiwen}), provide benchmark data against which molecular theories can be verified. In particular, one can examine if theories accurately capture the subtle changes that occur on the molecular scale, as chains begin to interact and interpenetrate with each other with increasing concentration. The aim of this paper is to carry out simulations with a recently developed multi-chain Brownian dynamics algorithm~\cite{JainPRE,JainCES}, which incorporates hydrodynamic and excluded volume interactions in order to compare predictions with experimental observations. Additionally, the technique of successive fine-graining~\cite{Sunthar,PrabhakarSFG} is used to obtain predictions that are independent of model parameters.

Over the past two decades, DNA (and in particular, $\lambda$-phage DNA) has been used as model polymer to carry out a number of investigations into single molecule dynamics. The advantage of DNA lies in the monodispersity of the solutions, and the ease with which the molecules can be stained with a dye for visual observation~\cite{Pecora}. For instance, in dilute solutions, single molecule studies of DNA have been used to examine the stretching dynamics of DNA molecules in extensional flows~\cite{Perkins,Smith&Chu}, stretching and tumbling dynamics in shear flows~\cite{Leduc,SmithShear}, dynamics in mixed shear and extensional flows~\cite{HazenMixed}, direct measurements of diffusion coefficients~\cite{Wirtz,RobertsonDiff} and relaxation times~\cite{PerkinsRelax}, and to establish the existence of coil-stretch hysteresis~\cite{CharlesCoilStretch}. In concentrated solutions, single molecule studies have established the validity of the reptation hypothesis~\cite{PerkinCon} and of scaling theories for the  molecular weight dependence of diffusion coefficients~\cite{SmithDif}. Compared to the wealth of experimental information on single molecule dynamics in the dilute and concentrated regimes, there is comparatively little  information on the behavior of macromolecules in the semidilute regime, both under equilibrium and non-equilibrium conditions. The classic early work of Chu and co-workers~\cite{Hur,BabcockPRL} was the first attempt to relate macroscopic rheological behavior to microscopic dynamics in shear flows. Steinberg and co-workers have measured the longest relaxation times for semidilute solutions of T4 DNA by carrying out stretch relaxation experiments~\cite{Steinberg}. More recently, Bausch and co-workers~\cite{Harasim,Huber} have correlated the dynamics of semiflexible polymers in semidilute solutions to the measured dependence of viscosity on shear rate. To our knowledge, there appear to be no measurements of single molecule dynamics in extensional flows of unentangled semidilute solutions, prior to the recent work of Hsiao \etal~\cite{Kaiwen}. It is also worth noting that experiments on single molecule behavior in extensional flows of dilute solutions have either separately examined the unravelling of chains from the coiled to the stretched state~\cite{Perkins,Smith&Chu,CharlesCoilStretch}, or the relaxation from the stretched to the coiled state~\cite{PerkinsRelax,CharlesCoilStretch}. The experiments of Hsiao \etal~\cite{Kaiwen} are unique in that they document the response of single chains to step-strain deformation followed by cessation of flow, both in the dilute and semidilute regime, and provide an opportunity to validate simulation predictions of chain stretch and relaxation in a single experiment.

In the case of dilute polymer solutions undergoing extensional flow, several studies have shown that it is necessary to incorporate the finite extensibility of chains, and the presence of hydrodynamic and excluded volume interactions into molecular theories in order to obtain an accurate prediction of experimental measurements~\cite{Shaqfeh,LarsonReview,LarsonExt,Jendrejack,Hsieh,CharlesExt,PrabhakarSFG}. In addition to having to choose the level of coarse-graining through a choice of the number of beads in a bead-spring chain, $N_b$, the incorporation of these phenomena entails the choice of parameters associated with each of them when carrying out simulations. Thus a choice needs to be made for the values of the nondimensional finite extensibility parameter, $b$, the nondimensional bead radius, $h^*$, which is a measure of the strength of hydrodynamic interactions, and the nondimensional excluded volume parameter, $z^*$, which is a measure of the difference between the solution temperature and the theta temperature. Prakash and co-workers~\cite{Sunthar,PrabhakarSFG,Pham} have shown that by using the method of successive fine-graining, predictions can be obtained that are independent of the choice of parameters in the model. The successive fine-graining technique is a specific protocol by which simulation data acquired for bead-spring chains with increasing values of $N_b$, is extrapolated to the number of Kuhn steps $N_k$ in the polymer chain being simulated. It essentially exploits the universal behavior observed in  solutions of long chain polymers, to obtain parameter free simulation predictions. In dilute solutions, the use of successive fine-graining has been shown to lead to quantitatively accurate predictions of the conformational evolution of $\lambda$-phage DNA in cross-slot cells~\cite{Sunthar} and the extensional viscosity of both DNA~\cite{SuntharVis} and polystyrene solutions~\cite{PrabhakarSFG,saadat2015molecular} in uniaxial extensional flows. The aim of the present paper is to use the successive fine-graining technique to predict the conformational evolution of DNA molecules in unentangled semidilute solutions when subjected to step-strain deformation followed by cessation of flow, and to verify if accurate predictions of the experimental measurements of Hsiao \etal~\cite{Kaiwen} can be obtained.

Several different mesoscopic simulation techniques have been developed over the past decade for describing the dynamics of unentangled semidilute polymer solutions which take into account the presence of intra and intermolecular long-range hydrodynamic interactions~\cite{Ahlrichs,BurkhardLadd,Huang,Stoltz,JainPRE,SaadatSemi}. By implementing the Kraynik-Reinelt periodic boundary conditions for mixed flows~\cite{Kraynik,Hunt}, Prakash and co-workers~\cite{JainCES} have recently developed an optimized multi-particle Brownian dynamics algorithm that can simulate arbitrary planar mixed shear and extensional flows of polymer solutions at finite concentrations. This algorithm is used in the present work to implement the successive fine-graining technique in the context of planar extensional flows.

The structure of the paper is as follows. In section~\ref{numdetails} and in the supplementary material, the governing equations for a bead-spring chain model are given along with the definitions of various observable quantities. In section~\ref{SFG}, a brief overview of the successive fine-graining technique is presented. A detailed comparison of simulation predictions with the experimental observations of~\citet{Kaiwen}, in dilute and in semidilute solutions, is presented in section~\ref{Results}. In particular, we carry out a qualitative comparison of the probability distribution of fractional stretch in planar extensional flows, and a quantitative comparison of the conformational evolution of individual chains subjected to a step-strain deformation followed by cessation of flow. A discussion of the reasons why the successive fine graining scheme may be expected to work, and the conditions under which it breaks down are discussed in section~\ref{blobology}. Finally, in section~\ref{conclu}, we summarise the principal conclusions of this work. 


\section{\label{numdetails}Bead-spring chain model of DNA}
A coarse-grained bead-spring chain model is used to represent DNA molecules, with each chain consisting of a sequence of $N_b$ beads (which act as centers of hydrodynamic resistance) connected by $N_{b} - 1$ massless wormlike chain (WLC) springs that represent the entropic force between two adjacent beads. A semidilute solution of DNA molecules is obtained by immersing an ensemble of $N_c$ such bead-spring chains in an incompressible Newtonian solvent. The bulk monomer concentration of the solution is defined by $c = {N}/{V}$, where $N = N_{b} \times N_c$ is total number of beads per cubic simulation cell of edge length $L_\text{sim}$, and $V = L_\text{sim}^3$, is the volume of each periodic cell. As detailed in the supplementary material, the inter and intramolecular hydrodynamic interactions between the beads are modeled using the Rotne-Prager-Yamakawa (RPY) tensor, while bead overlap is prevented by using a pairwise repulsive narrow Gaussian excluded volume potential. For the purpose of non-dimensionalising length and time units, a length scale $l_{H} = \sqrt{k_{B}T/H}$ and a time scale $\lambda_{H}=\zeta/4H$  are used respectively, where $T$ is the temperature, $H$ is the spring constant, $k_{B}$ is the Boltzmann constant, and $\zeta$ is the hydrodynamic friction coefficient associated with a bead. Within this framework, the time evolution of the dimensionless position, $\Vector{r}^*_{\nu} (t^*)$ of a typical bead $\nu$, is governed by a stochastic differential equation, which can be numerically integrated with the help of Brownian dynamics simulations. The specifics of the integration scheme, along with details of the simulation protocol, and the particular forms of the spring force and the hydrodynamic interaction tensor used here, are given in the supplementary material.  

DNA solutions used in rheological measurements are typically buffered aqueous solutions with an excess concentration of sodium salt, which has been established to be well above the threshold for observing charge-screening effects (see Appendix~B of~\citet{pan2014universal}). Consequently, DNA molecules are expected to behave identically to neutral molecules in good solvents that lie in the crossover regime between $\theta$ solutions and athermal solvents, with the solvent quality described by the variable~\cite{pan2014universal,pan2014viscosity},
\begin{equation}
\label{SolQua}
z=k \left( 1 - \frac{T_{\theta}}{T}\right)\sqrt{M}
\end{equation}
where $M$ is the molecular weight, $T_{\theta}$ is the theta temperature, and $k$ is a polymer-solvent chemistry dependent constant. Recently, Pan \etal\ have estimated that $T_{\theta} \approx 15\degC$ for the DNA solutions that are typically used in rheological experiments, and have also determined the value of the constant $k$~\cite{pan2014universal,pan2014viscosity}. In particular, they have tabulated the value of $z$ as a function of temperature and molecular weight for a wide variety of DNA fragments. Based on their calculations, a solution of $\lambda$-phage DNA  is estimated to have a solvent quality $z\approx 0.7$ at 22\degC\ (the temperature at which the experiments reported in Hsiao \etal~\cite{Kaiwen} have been carried out). Interestingly, Sunthar \etal~\cite{Sunthar,SuntharVis} found that using $z=1$ (rather than $z=0$ or $z=3$) in their dilute solution simulations gave the best agreement between predictions and the experimental measurements of~\citet{Perkins}. At equilibrium, experiments and simulations show that for  $z=0.7$, $R_g = 1.23 \, R_g^\theta$, while at $z=1.0, R_g = 1.29 \, R_g^\theta$ (these  estimates can be obtained from Eqn.~23 of Ref.~\cite{Kumar} which gives a formula for $R_g/R_g^\theta$ as a function of $z$ that fits both experimental and simulation data). The difference in equilibrium chain swelling between the two values of $z$ is consequently less than 5\%, which is not significant in comparison to experimental and simulation error bars. Anticipating, therefore, that the difference between results for these two values of $z$ will not be significant, we have used a value of $z=1$ in all our simulations. However, in order to ensure that this is in fact the case, we have validated this assumption by carrying out simulations with $z=0.7$, at one value of the Weissenberg number ($W\!i = 2.6$).  The results of the study, which are presented in the supplementary material, show that indeed this assumption is justified. 

The solvent quality can be conveniently controlled in simulations with the help of the narrow Gaussian potential~\cite{RaviExclu,Kumar}
\begin{equation}
\label{exvol}
E(\Vector{r}^*_{\nu \mu}) = \left(\frac{z^*}{{d^*}^3}\right) \exp \left\lbrace -\frac{1}{2}\frac{{\Vector{r}^*_{\nu \mu}}^2}{{d^*}^2} \right\rbrace 
\end{equation}
which determines the force due to excluded volume interactions between any two beads $\mu$ and $\nu$. Here, $z^*$ is the strength of the excluded volume interactions, and $d^*$ is the range of the interaction. A mapping between experiments and simulations is achieved by setting $z=z^* \sqrt{N_b}$, with $z^*$ being a measure of the departure from the $\theta$-temperature, and $N_b$ being proportional to the molecular weight~\cite{Kumar,sunthar2006dynamic}. As a result, for any choice of $N_b$, $z^*$ is chosen to be equal to $z/\sqrt{N_b}$ such that the simulations correspond to the given experimental value of $z$. For reasons elaborated in Refs.~\cite{prakash2001influence,Kumar} in the context of dilute polymer solutions, the parameter $d^*$ is irrelevant for sufficiently long chains, and is typically calculated by the expression $d^* = K {z^*}^{1/5}$, with $K$ being an arbitrary constant. It is worth noting that in order to establish that simulation predictions obtained with the successive fine-graining protocol are truly parameter free, it is necessary to demonstrate independence from the choice of the constant $K$ in addition to the other model parameters discussed earlier. In the present instance, the influence of $K$ on simulation predictions is examined in the supplementary material, and shown to be irrelevant as expected.

A majority of the experimental measurements by~\citet{Kaiwen} in the semidilute regime have been carried out at the scaled concentration $c/c^* = 1$, where $c^*$ is the overlap concentration, which is defined here by the expression, $c^{*} = N_{b}/\left[(4\pi/3)(R^{0}_{g})^{3}\right]$, with $R^{0}_{g}$ being the radius of gyration for an isolated chain at equilibrium. The value of $c/c^*$ is calculated for each simulation reported here by computing $R^{0}_{g}$ a~priori from single-chain BD simulations at equilibrium, for the relevant set of parameter values. 

The velocity gradient tensor for planar extensional flows is given by~\cite{BirdVol1}
\begin{equation}
\label{tensor}
(\bm{\nabla}\bm{v^*})_{\text{PEF}} =  \begin{pmatrix}
\dot{\epsilon}^* & 0 & 0 \\
0 & -\dot{\epsilon}^* & 0 \\
0 & 0 & 0 \\
\end{pmatrix}
\end{equation}
where $\dot{\epsilon}^*$ is the elongation rate. Planar extensional flows are generally difficult to study by computer simulations, since fluid elements are exponentially stretched in one direction and contracted in the perpendicular direction. This leads to a very short window of time to observe the dynamics of single molecules since the dimensions of the simulation box rapidly become of order of intermolecular distance. This difficulty can be resolved by the implementation of Kraynik-Reinelt periodic boundary conditions~\cite{Kraynik,Todd,Baranyai}. As mentioned earlier, Jain \etal~\cite{JainCES} have implemented these boundary conditions for BD simulations in the context of arbitrary planar mixed flows, and this algorithm has been adopted here.         

Simulation predictions are compared with the experimentally measured \emph{stretch} of molecules, when a semidilute solution is subjected to a step-strain deformation in a planar extensional flow. The stretch of a fluorescently dyed DNA molecule, measured in a cross-slot cell, is the projected extent of the molecule in the flow direction. For a bead-spring chain model, this is calculated from,
\begin{equation}
\label{stretch}
X_{\text{max}}^* \equiv \text{max}_{\mu,\,\nu} | {r^*_\mu}^x - {r^*_\nu}^x |
\end{equation}
where ${r^*_\mu}^x$ is the $x$-component of the vector $\Vector{r}^*_\mu$ of bead $\mu$, with $x$ being the direction of flow. The mean stretch can be obtained from the bead positions in an ensemble of chain configurations from the ensemble average,
\begin{equation}
\label{stretchmean}
\bar{X}^* = \avg{X^*_\text{max}}
\end{equation}
The equilibrium mean stretch is denoted by $\bar{X}^*_\text{eq}$. Experimental measurements of stretch are typically reported in terms of the nondimensional ratio $\bar{X}/L$, where $L$ is the contour length of stained $\lambda$-phage DNA molecules, typically assumed to be equal to 22 $\mu$m. However, we often find it convenient to additionally use the expansion ratio,
\begin{equation}
\label{ExpRatio}
E = \frac{\bar{X}^*}{\bar{X}^*_\text{eq}}
\end{equation}
in simulations.

The longest relaxation time $\lambda_1$ is measured experimentally by fitting the terminal 30$\%$ of the stretch of a molecule, as it relaxes from a highly extended state, with a single exponential decay~\cite{Kaiwen}. In simulations, the longest dimensionless relaxation time $\lambda_1^*= \lambda_1/ \lambda_H$, for any bead-spring chain with $N_b$ beads, is obtained by initially stretching each chain to nearly 90$\%$ of its fully extended state, and letting it relax to equilibrium. Details of this procedure are presented in the supplementary material. 

\section{Successive fine-graining}
\label{SFG}
The successive fine-graining technique exploits the universal behavior of polymer solutions to obtain property predictions that are independent of the choice of model parameters. At equilibrium, this technique has been widely used to obtain universal predictions from analytical theories and molecular simulations~\cite{zimm1980chain,Kumar,garcia1984monte,freire1986monte,garcia1991monte,sunthar2006dynamic,pan2014viscosity,JainPRL}. Essentially, data is accumulated for finite chains, and subsequently extrapolated to the long chain limit, $N_b\rightarrow \infty$, where the self-similar character of polymer chains is captured. Extrapolation to the long chain limit has also been used to obtain universal predictions in shear flow, where the finiteness of chain length is not relevant for sufficiently long chains at typically measured shear rates~\cite{kroger2000variance,ottinger1987generalized,ottinger1989gaussian,prakash1997universal,prakash2002rouse,kumar2004universal}. In extensional flows, however, where at high extension rates chains are nearly fully stretched, the finiteness of chain length plays a crucial role in determining the solution's properties. Even under these circumstances, provided the flow has not `penetrated' below the Pincus blob length scale, universal behavior is still observed~\cite{Sunthar,somani2010effect} (see also discussion in section~\ref{blobology}). Prakash and co-workers have modified the successive fine-graining technique for infinitely long chains, by making it applicable under conditions where it is important to account for the finite length of a chain~\cite{Sunthar,PrabhakarSFG,Pham,bosko2011universal}. While at its core, the modification consists of changing the extrapolation limit from $N_b\rightarrow \infty$ to $(N_{b} - 1)\rightarrow N_k$, where $N_k$ is the number of Kuhn steps in the underlying chain, the details of the method are more subtle and complex. Sunthar and Prakash have discussed the procedure in great detail in Ref.~\cite{Sunthar}. For the sake of completeness, and since it is used in the context of semidilute solutions here for the first time, we briefly motivate and explain the salient features of the technique below.

An example of a universal equilibrium property for dilute polymer solutions under $\theta$ conditions is the Flory-Fox constant $U_{\eta R}^\theta$, defined by~\cite{Rubinstein}, 
\begin{equation}
\label{FloryFoxConstant}
U_{\eta R}^\theta = \frac{[\eta]_{\theta}M}{\left({4 \pi}/{3} \right) {\left(R_{g}^{\theta}\right)}^3 N_{A}}
\end{equation}
where $R_{g}^{\theta}$ is the radius of gyration, $[\eta]_\theta$ is the zero shear rate intrinsic viscosity, and $N_A$ is Avagadro's constant. It is a surprising experimental observation that $U_{\eta R}^\theta$ attains its universal value of $1.49 \pm 0.06$ for a wide range of polymer-solvent systems~\cite{miyaki1980flory}, for molecular weights as low as $M = 50,000$ g/mol~\cite{krigbaum1,krigbaum1953molecular}. As a result, it is clear that the intrinsic viscosity at the $\theta$ temperature for a majority of dilute solutions of linear flexible polymers can be calculated once the radius of gyration of the polymer under $\theta$ conditions is known. For polymer solutions in the crossover region between $\theta$ and very good solvents, an additional variable, namely the solvent quality parameter $z$ is required to describe universal behavior. For instance, for a number of different polymer-solvent systems, the ratio,
\begin{equation}
\label{RatioViscosity}
\alpha_\eta (T,M) = \left( \frac{[\eta]}{[\eta]_{\theta}}\right)^{1/3}
\end{equation}
measured at different temperatures and molecular weights, is found to collapse onto a master plot, when plotted as a function of $z$~\cite{Tominaga20021381,pan2014viscosity}. Since,
\begin{equation}
\label{Eta}
[\eta](T, M)  = [\eta]_{\theta} \, \alpha_\eta^{3} 
              = U_{\eta R}^{\theta} \left( \frac{N_{A}}{M}\right) \left( \frac{4 \pi}{3} R_{g}^{\theta}\right)^{3} \left[\alpha_{\eta} (z) \right]^{3}
\end{equation}
it is clear that a knowledge of $R_g^{\theta}$, and the universal properties $U_{\eta R}^{\theta} $ and $\alpha_\eta(z)$, enables the determination of the intrinsic viscosity of any dilute linear polymer-solvent system in the crossover regime. A similar argument can be made for any other static or dynamic property of a dilute polymer solution, $\phi (T, M)$. Essentially, provided one knows a suitably defined universal ratio $U_{\phi R}^{\theta}$ under $\theta$ conditions, and the universal crossover swelling function $\alpha_\phi (z) = \phi (z)/\phi_\theta$, the property $\phi$ can be determined for the solution at any temperature and polymer molecular weight, given $R_g^{\theta}$ and $z$. This is the basic content of the two-parameter theory~\cite{yamakawa1971modern}, which states that all static and dynamic properties of a dilute solution of linear flexible polymers can be determined once $R_g^{\theta}$ and $z$ are known.

Bead-spring chain models with Hookean springs need three parameters $\left\{N_{b}, h^*, z^*\right\}$, to be specified, when nondimensionalized with the length scale $l_H$, and time scale $\lambda_H$. While the strength of hydrodynamic interactions is specified by the draining parameter~\cite{zimm1956dynamics,ottinger1989renormalization}, $h = h^{*} \sqrt{N_b}$, the strength of excluded volume interactions~\cite{schafer2012excluded,prakash2001rouse} is determined by $z = z^* \sqrt{N_b}$. Typically, the parameters $h^*$ and $z$ are kept constant when implementing the successive fine-graining procedure of extrapolating finite chain data to the long chain limit, $N_b \rightarrow \infty$~\cite{zimm1980chain,Kumar,freire1986monte,garcia1991monte,sunthar2006dynamic,pan2014viscosity,JainPRL,kroger2000variance,ottinger1987generalized,ottinger1989gaussian,prakash1997universal,prakash2002rouse}. This implies that universal property predictions at equilibrium and in shear flow are obtained in the non-draining limit $h \rightarrow \infty$ (independent of the particular choice made for $h^*$), and at a specific location in the crossover regime specified by the solvent quality $z$. 

The modified successive fine-graining procedure for polymer solutions in extensional flows~\cite{Sunthar,PrabhakarSFG} also leads to universal predictions in the limit of large $h$ and constant $z$. However, the use of finitely extensible springs in place of Hookean springs, in order to account for finite chain length, leads to significant changes in the implementation of the procedure. 

When subjected to extensional flow, a dilute polymer solution in the crossover regime is characterized by the following set of variables: $\left\{ R_{g}^{\theta}, z, L, W\!i, \epsilon\right\}$. Here, $L$ is the finite contour length of the chain, $W\!i = \lambda_{1} \, \dot{\epsilon} $ is the Weissenberg  number, with $\dot{\epsilon}$ being the extension rate, and $\epsilon = \dot{\epsilon} \, t $ the Hencky strain, which measures the extent of deformation from the onset of flow. The protocol for successive fine-graining of finite chains described briefly below, ensures that universal property predictions are obtained for this set of prescribed experimental variables.

The maximum number of conformational degrees of freedom for a finite chain is the number of Kuhn steps, $N_k$. Extrapolation of finite chain data can consequently only be carried out to the limit $(N_{b} - 1) \rightarrow N_k$. The number of Kuhn steps in a flexible linear chain can be determined from the expression, 
\begin{equation}
\label{KuhnEq}
N_k = \frac{L^2}{6(R_g^\theta)^2}
\end{equation}
While the $\theta$ temperature for DNA in aqueous solutions with excess sodium salt (typically used for cross slot flow measurements), has been shown to be roughly $15^{\circ}{\rm C}$ by Pan \etal~\cite{pan2014universal}, there does not yet seem to be an accurate measurement of $R_g^\theta$. In the absence of information on $R_g^\theta$, $N_k$ can also be found from the expression $N_k = L/(2\lambda_p)$, where $\lambda_p$ is the persistence length. In Appendix B of Ref~\cite{pan2014universal}, Pan \etal~have reported measurements of $\lambda_p$ by various authors, using a variety of different techniques, to be roughly 50 nm in the presence of excess sodium salt. As a result, using a contour length of 16 $\mu$m, suggests $N_k = 160$. On the other hand, staining with YOYO-1 dye is known to increase the contour length~\cite{Perkins,Smith&Chu}. The recent experiments by the Doyle group~\cite{kundukad2014effect} suggest that the contour length is increased by 38$\%$ at full saturation of one YOYO-1 per four base pairs of DNA. For $\lambda$-phage DNA, this implies a stained contour length of 22 $\mu$m, in agreement with earlier estimates~\cite{Perkins,Smith&Chu}. If the persistence length remains unchanged subsequent to the intercalation by the dye, as suggested in Ref.~\cite{kundukad2014effect}, then the number of Kuhn steps would be roughly $N_k = 220$. Sunthar and Prakash~\cite{Sunthar} have argued that results of the successive fine-graining procedure are insensitive to a choice of $N_k$ in the range 150-300, and have used $N_k = 200$ in their simulations of dilute $\lambda$-phage DNA solutions subjected to extensional flow. It is worth noting that since results are extrapolated to the limit $\sqrt{1/N_k}$, this range of $N_k$ values implies extrapolating finite chain data to either 0.08 or 0.06. While we have not carried out extensive studies to investigate the influence of the choice of $N_k$ for semidilute solutions, we have adopted the value $N_k = 200$ in the current simulations based on these arguments.

The centrality of the finiteness of chain length is maintained in the successive fine-graining procedure by ensuring that at every level of coarse-graining, the fully stretched length of the bead-spring chain is identical to the contour length of the polymer being modelled. As a consequence, for any choice of the number of beads $N_b$
\begin{equation}
\label{contourlength}
 L  = (N_b - 1) \sqrt{b} \, l_H
\end{equation}
In order to be consistent with the equilibrium properties of the polymer, it is also required that the radius of gyration of the bead-spring chain under $\theta$ conditions remains unchanged with fine-graining. 

Defining the dimensionless mean square length of a single finitely extensible spring in the bead-spring chain, $\chi^2 (b)$, by
\begin{equation}
\label{AvgQ}
\chi^2 (b) = \frac{\avg{Q^2}}{3\, l_H^2}
\end{equation}
where $\avg{Q^2}$ is the dimensional mean-square end-to-end vector of a single spring, it is straight forward to show that~\cite{BirdVol2,Sunthar}
\begin{equation}
\label{RgExpre}
\left(R_g^\theta \right)^2 = \chi^2 (b) \, \frac{(N_b^2 - 1)}{2 N_b} \, l_H^2 
\end{equation}
Evaluating the ratio $L^2/ \left(R_g^\theta\right)^2$ from Eqs.~\ref{contourlength} and~\ref{RgExpre}, and using the definition of $N_k$ in Eq.~\ref{KuhnEq} implies,
\begin{equation}
\label{RelbetbandChi}
\frac{b}{\chi^2(b)}=\frac{3(N_b + 1)}{N_b (N_b - 1)}N_k
\end{equation}
Sunthar and Prakash~\cite{Sunthar} have shown that for wormlike chains,
\begin{equation}
\label{RelbandchiWLC}
\frac{\chi^2(b)}{b}=\frac{1}{3} \, \frac{\int_{0}^{1}dq^* \, {q^*}^4 \, e^{-\phi_\text{c}^* (b,q^*)}}{\int_{0}^{1}dq^* \, {q^*}^2 \, e^{-\phi_\text{c}^* (b,q^*)}}
\end{equation}
where $\phi_\text{c}^*$ is the nondimensional spring potential,
\begin{equation}
\label{springpot}
\phi_\text{c}^*(b, q^*) = \frac{b}{6} \, \left[ 2 \, {q^*}^2 + \frac{1}{1-q^*} - q^*\right]
\end{equation}
Equations~\ref{RelbetbandChi}, \ref{RelbandchiWLC} and~\ref{springpot} enable the determination of the finite extensibility parameter $b$, and the nondimensional mean square length of a single spring $\chi^2 (b)$, for any choice of $N_b$ and $N_k$. A simple and efficient procedure for calculating $b$ and $\chi^2 (b)$ has been described in Ref.~\cite{Sunthar}.

The quantity $\chi^2 (b)$ also plays an important role in the treatment of hydrodynamic and excluded volume interactions in the successive fine-graining procedure. For a bead-spring chain with finitely extensible springs, the draining parameter can be shown to be given by the expression~\cite{Sunthar}, $h = \tilde{h}^*\sqrt{N_b}$, where
\begin{equation}
\label{relhstar}
\tilde{h}^* = \frac{h^*}{\chi(b)}
\end{equation}
while the solvent quality can be shown to be given by~\cite{Sunthar}, $z = \tilde{z}^*\sqrt{N_b}$, where
\begin{equation}
\label{relzstar}
\tilde{z}^* = \frac{z^*}{\left[\chi(b) \right]^3}
\end{equation}
Note that $\chi (b) \to 1$ in the limit $N_b \rightarrow \infty$. When carrying out the successive fine-graining procedure for infinite chains, as mentioned earlier, the parameter $h^*$ is held constant as $N_b \rightarrow \infty$, while $z^*$ is calculated from $z^* = z/\sqrt{N_b}$ at each level of fine-graining. On the other hand, during the successive fine-graining procedure for finitely extensible bead-spring chains, $\tilde{h}^*$ is held constant at each level of fine-graining, which implies, $h^* = \tilde{h}^* \, \chi(b)$, and $z^*$ is calculated from the expression, $z^* = \left(z/\sqrt{N_b} \right) \left[ \chi (b)\right]^3$. Sunthar and Prakash~\cite{Sunthar} and Pham \etal~\cite{Pham} have shown that at equilibrium (where $W\!i$ and $\epsilon$ are not relevant variables), extrapolation of finite chain data to the limit $(N_b - 1) \rightarrow N_k$, using this procedure, leads to property predictions that are in quantitative agreement with known results for bead-rod chains with $N_k$ rods. Additionally, Pham \etal~\cite{Pham} established the validity of the successive fine-graining procedure in steady shear flow by comparing bead-spring chain results with the results of a bead-rod model and a stiff FENE--Fraenkel spring model, both in the absence and presence of hydrodynamic and excluded volume interactions.

For a polymer solution subjected to extensional flow, if comparison of simulation predictions is being made with experimental data at particular values of $W\!i$ and $\epsilon$, the successive fine-graining procedure ensures that at each level of coarse-graining, simulations are carried out at the same values of $W\!i$ and $\epsilon$. This is achieved by the following series of steps. (i) For any choice of $N_b$, chains are stretched to nearly $90\%$ of their fully stretched state and allowed to relax. The longest relaxation time $\lambda_1^*$ (at that value of $N_b$) is then found by fitting a single exponential decay to the terminal $30\%$ of the mean stretch, as described in the supplementary material. (ii) The extension rate $\dot{\epsilon^*}$ used for simulation of chains with $N_b$ beads is then found from the expression, $\dot{\epsilon^*} = W\!i/\lambda_1^*$, where $W\!i$ is the experimental Weissenberg number. (iii) Once $\dot{\epsilon^*}$ is known for any $N_b$, simulations are carried out until a nondimensional time $t^*$, such that $\dot{\epsilon^*} \, t^* = \epsilon$. By maintaining $W\!i$ and $\epsilon$ identical to experimental values at each level of fine-graining in this manner, we ensure that the extrapolated results in the limit $(N_b - 1) \rightarrow N_k$ are also at the specified experimental values.

To date, the successive fine-graining procedure for \emph{finite} chains has only been used in the context of \emph{dilute} polymer solutions~\cite{Sunthar,PrabhakarSFG,Pham,saadat2015molecular}. Recently Jain \etal~\cite{JainPRL} have extrapolated finite chain data  to the \emph{long chain} limit in the \emph{semidilute} regime, to obtain universal predictions of the ratio of semidilute to dilute single chain diffusion coefficients at various values of concentration. In the present paper, we use the successive fine-graining procedure for finite chains to compare simulation predictions for extensional flows of semidilute solutions with the experimental measurements of Hsiao \etal~\cite{Kaiwen}.

\section{Results and discussion}
\label{Results}  

\begin{figure}[t]
\centering
\includegraphics[width=8.5cm,height=!]{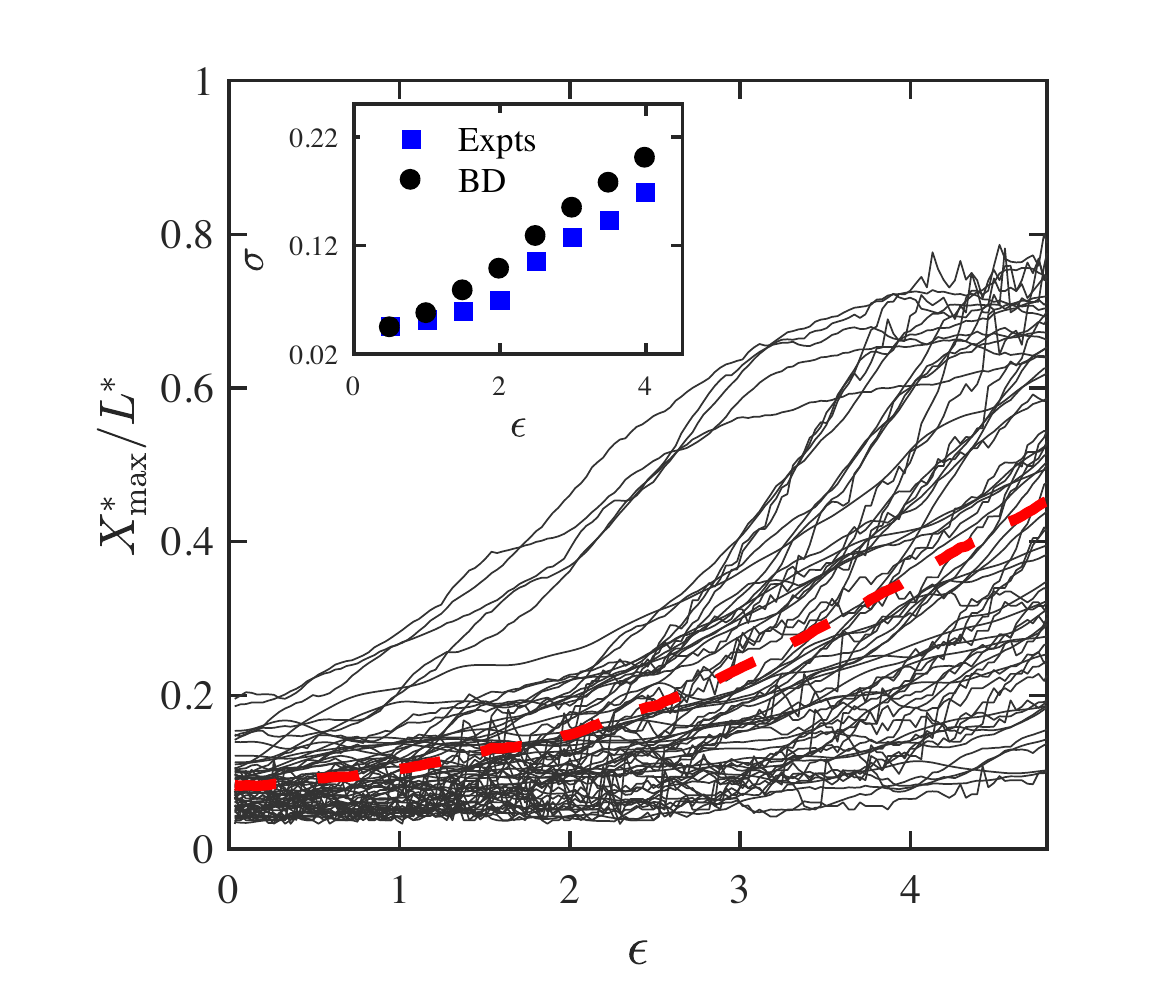}
\caption{\footnotesize Evidence of molecular individualism during stretching. The black curves are individual trajectories of 67 chains, while the dashed red curve is the ensemble average over the chains ($\bar X/L$). The inset compares the standard deviation in the experimental and simulation stretch data as a function of strain. Parameter values for the simulation are:  $N_{b} = 45,\, c/c^* = 1,\, z = 1, \,\tilde{h}^* = 0.19$,\, $ N_{k} = 200$ and $W\!i = 2.6$.}
\label{DifferTraj}
\end{figure}

\begin{figure}[!ht]
\centering
\includegraphics[width=8.5cm,height=!]{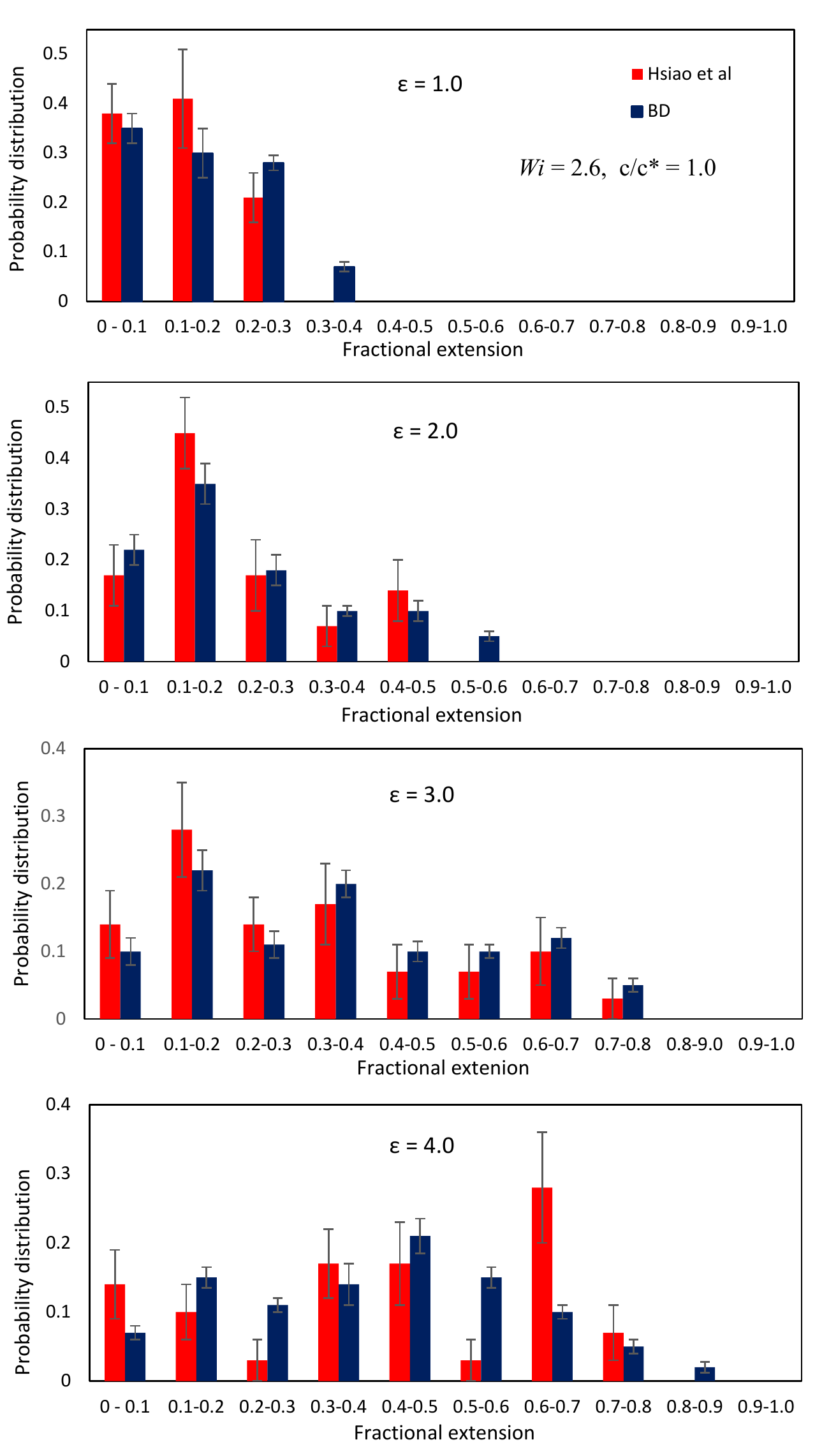}
\caption{\footnotesize Probability distribution of chain extension in a semidilute solution at $c/c^* = 1$. Distributions are shown for a range of accumulated strains $\epsilon$ at a Weissenberg number $W\!i = 2.6$. Red histograms are the experimental results of~\citet{Kaiwen}, while the blue histogram are the results of Brownian dynamics simulations with parameter values:  $N_{b} = 45, \, z = 1, \, \tilde{h}^* = 0.19$, and $N_{k} = 200$.}
\label{probdis}
\end{figure}

\begin{figure}[!ht]
\centering
\includegraphics[width=8.5cm,height=!]{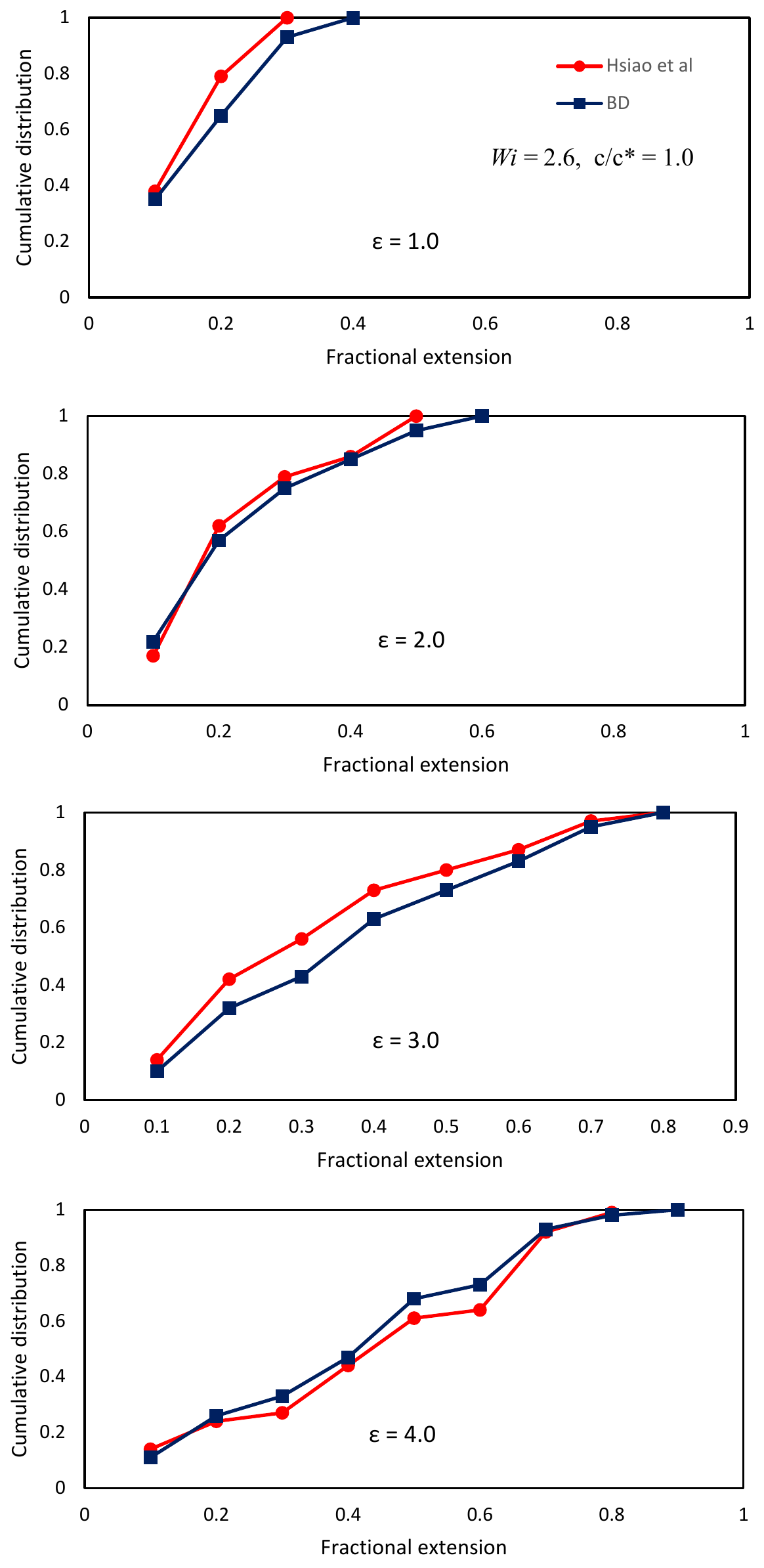}
\caption{\footnotesize Cumulative probability distribution of chain extension in a semidilute solution at $c/c^* = 1$. Distributions are shown for a range of accumulated strains $\epsilon$ at a Weissenberg number $W\!i = 2.6$. Red curves are the experimental results of~\citet{Kaiwen}, while the blue curves are the results of Brownian dynamics simulations with parameter values:  $N_{b} = 45, \, z = 1, \, \tilde{h}^* = 0.19$, and $N_{k} = 200$.}
\label{cumuprobdis}
\end{figure}

A striking early observation of single molecule experiments in dilute solutions~\cite{Smith&Chu} was the enormous variability in the transient stretching dynamics of the different molecules, a phenomena characterised by de Gennes as `molecular individualism'~\cite{deGennesMol}. \citet{Kaiwen} have observed a similarly wide distribution of configurations in their observation of individual molecular trajectories at $c/c^* = 1$, albeit with qualitatively different molecular conformations in semidilute solutions compared  to dilute solutions. Individual trajectories obtained by simulating 67 chains in the main simulation box (with $N_b = 45$ and parameter values reported in the figure caption) are displayed by the black curves in Fig.~\ref{DifferTraj}. The dashed red curve is the ensemble average over the chains. Clearly, wide variability in the manner in which chains unravel from the coiled to the stretched state is also observed in our simulations of extensional flow. The inset to Fig.~\ref{DifferTraj}, which compares the standard deviation in the experimental and simulation stretch data as a function of strain, reveals that the spread of stretch values is of similar magnitude in both cases.

A qualitative comparison of the probability distribution of chain extension observed in a simulation with $N_b = 45$, and the experiments of \citet{Kaiwen}, is shown in Fig.~\ref{probdis}. Essentially 50 simulations, each with 67 chains in the main simulation box, were carried out and the fractional extension $(X_\text{max}^*/L^*)$ for each of the chains was calculated at various values of $\epsilon$, and the results were binned as indicated in the figure. Here, $L^* = (N_b-1) \, \sqrt{b}$. The number of chains in each of the bins,  $0 \le (X_\text{max}^*/L^*) < 0.1$,  $0.1 \le (X_\text{max}^*/L^*) < 0.2$, etc., was divided by 3350 (the total number of chains in the sample), to obtain the probability distribution. Fig.~\ref{cumuprobdis} represents the fractional extension of the ensemble of chains as a cumulative distribution, and gives an alternative perspective of the same data. Note that the method of successive fine-graining has not been applied and the simulation results are at a single value of $N_b$. Nevertheless, a good qualitative agreement can be observed, with simulations reflecting the experimental observation of a broadening of the distributions as the accumulated strain increases, with the persistence of chains that remain partially unravelled even at high strains. There is greater  variability between the results of simulations and experiments at high fractional extensions and high strain. As will be clear in the subsequent discussion of the results of successive fine graining, it is essential to capture the many degrees of freedom in the real system in order to get close agreement between experimental and simulation results. 

\begin{figure}[t]
\centering
\includegraphics[width=8.5cm,height=!]{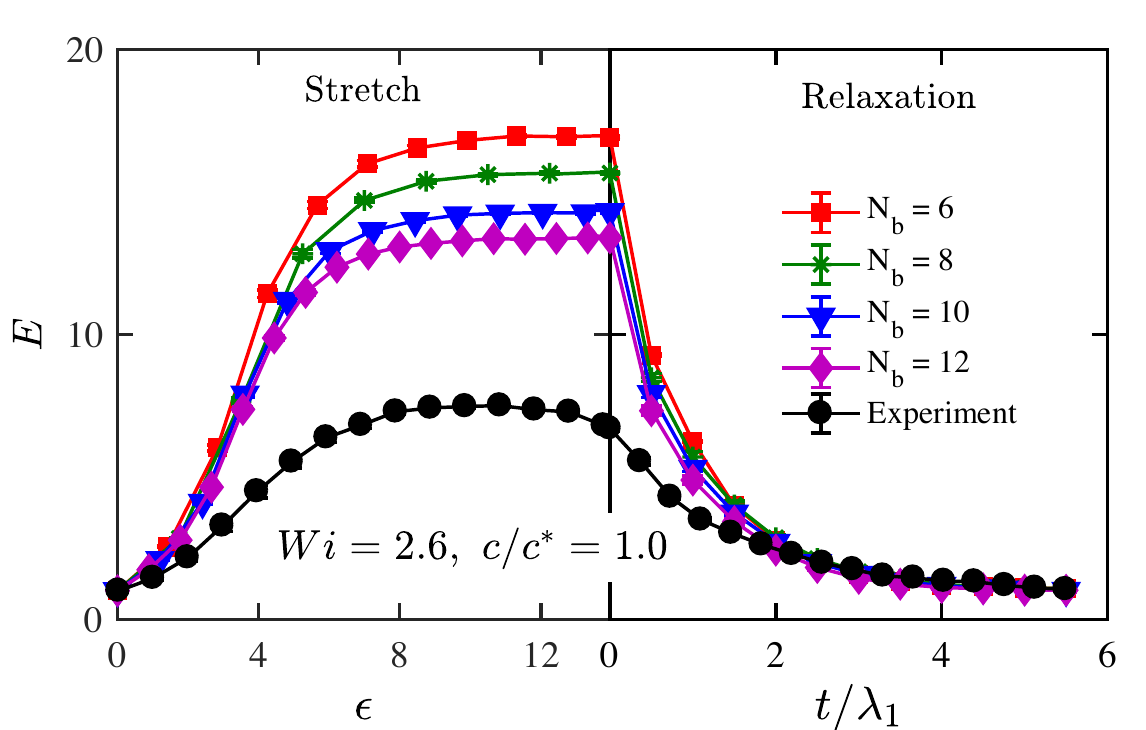}
\caption{\footnotesize Transient polymer stretch in a step strain experiment in planar extensional flow at $c/c^* = 1$ and $W\!i = 2.6$. The black line and symbols are experimental measurements of the ensemble average stretch ratio by~\citet{Kaiwen} and the various coloured lines and symbols are BD simulations at the various values of $N_b$ indicated in the legend. Common parameter values in all  the simulations are: $z = 1, \, \tilde{h}^* = 0.25$, and $ N_{k} = 200$. Values of $b$, $\chi(b)$, $h^*$, $z^*$, $\lambda_1^*$ and $\dot \epsilon^*$ used for each of the simulated values of $N_b$, are calculated as per the procedure described in section~\ref{SFG}.  }
\label{FullStretchRelaxation}
\end{figure}

As mentioned earlier, the unique character of the single molecule experiments of~\citet{Kaiwen} is the implementation of a step input on the strain rate $\dot \epsilon$, followed by the cessation of flow once the fluid has accumulated a Hencky strain of $\epsilon$. This enables the observation of the non-equilibrium stretching and relaxation dynamics in a single experiment.  Fig.~\ref{FullStretchRelaxation} compares the experimental measurements of the ensemble average stretch ratio $E$ by~\citet{Kaiwen} at  $c/c^* = 1$, and $W\!i = 2.6$, with BD simulations carried out at various values of $N_b$. The flow is maintained until $\epsilon = 13$, before being switched off, and the subsequent relaxation is observed for a period of time measured in terms of the nondimensional units, $t/\lambda_1$. The use of the stretch ratio and non-dimensional time as the axes enables a direct comparison of simulation and experiments. Clearly, the qualitative behaviour observed in experiments is captured in the simulations. The chains unravel from the coiled state and reach a steady-state value of stretch after about 8 Hencky strain units. While the curves for the different values of $N_b$ are quite different from each other in the stretch phase, they become more tightly bunched together as the chains relax towards their equilibrium coiled state. This is because all chains, regardless of their length, relax to a common value of $E=1$ at long times. In spite of the simulation predictions becoming closer to experimental measurements for increasing values of $N_b$, the significant quantitative difference between simulations and experiment at all values of $N_b$ reported in Fig.~\ref{FullStretchRelaxation}, points to the importance of capturing all the degrees of freedom of the polymer chain being simulated. This is precisely the purpose of successive fine-graining, which we carry out below. 

\begin{table*}[t]
\caption{\footnotesize Typical values of simulation parameters that arise at each level of coarse-graining when carrying out the successive fine-graining procedure for semidilute simulations, corresponding to the following set of experimental values:  $\{c/c^* = 1,\, z = 1, \, N_{k} = 200$ and $W\!i = 2.6\}$. The hydrodynamic interaction parameter was maintained constant at $\tilde{h}^* = 0.19$.}
\centering
\vskip10pt
\label{ParaValue}
\setlength{\tabcolsep}{10pt}
{\def\arraystretch{1.3}
\begin{tabular}{c  c  c  c  c  c c  c  c}
\hline \hline
$N_b$ & $b$ & $\chi (b)$ & $z^*$ & $h^*$ & $\bar{X}^*_\text{eq}$ & $\lambda^{*}_1$ & $\dot{\epsilon}^*$\\
\hline \hline
6 & 124.04 & 0.9413 & 0.3404 & 0.1788 & $2.127 \pm 0.002$  & 11.021 & 0.2359\\
8 & 82.652 & 0.9258 & 0.2805 & 0.1759 & $2.904 \pm 0.003$ & 17.826 &  0.1458\\
10 & 60.911 & 0.9114 & 0.2393 & 0.1731 & $3.455 \pm 0.002$ & 25.883 &  0.1004\\
12 & 47.609 & 0.8976 & 0.2087 & 0.1705 & $4.047 \pm 0.023$ & 35.104 & 0.0740 \\
\hline \hline
\end{tabular}
}
\vskip-5pt
\end{table*}

\begin{figure}[!ht]
\centering
\includegraphics[width=8.5cm,height=!]{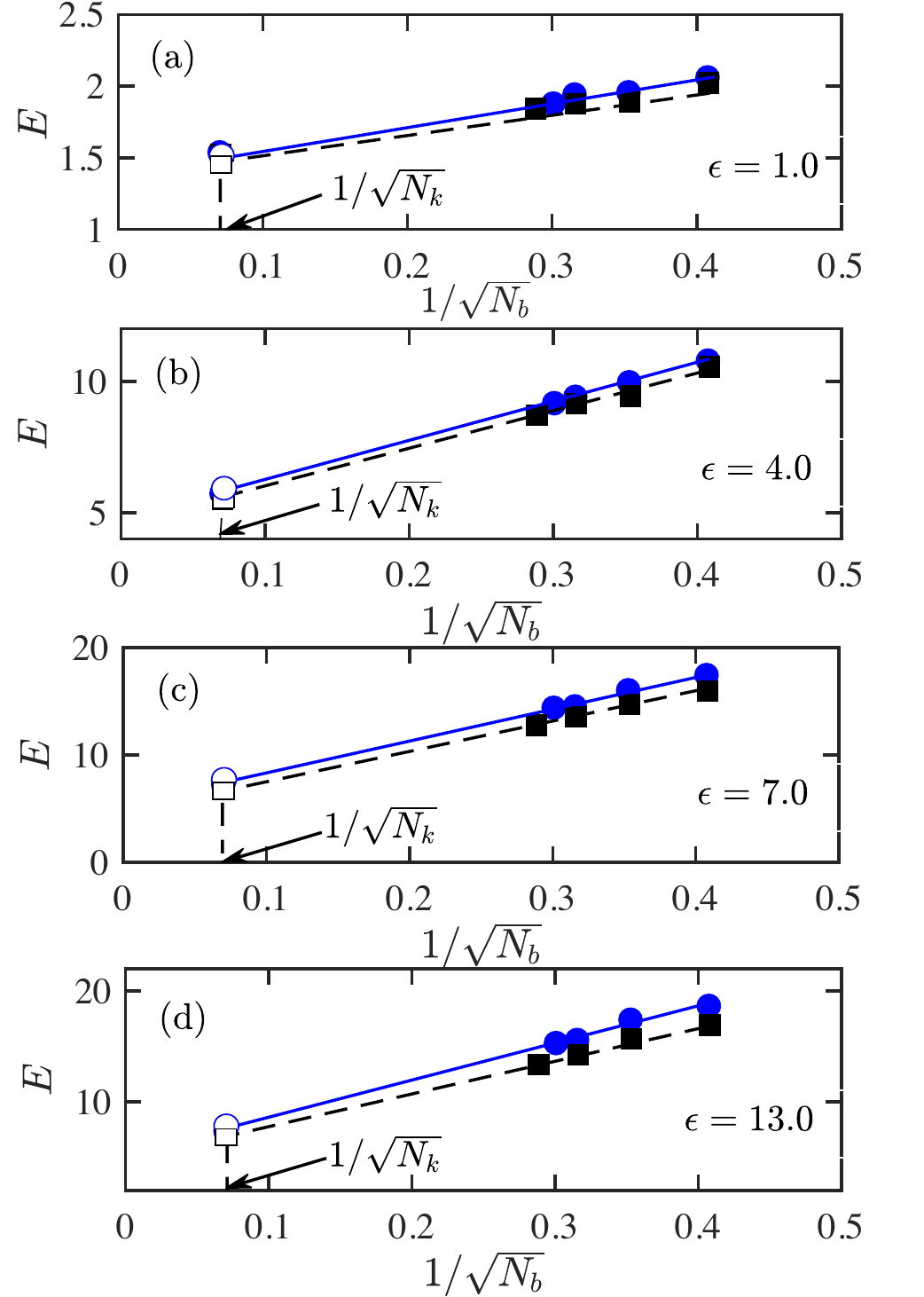}
\caption{ \footnotesize Illustration of the extrapolation procedure during the stretching phase [(a) $\epsilon = 1$, (b) $\epsilon = 4$, (c) $\epsilon = 7$ and~(d) $\epsilon = 13$], for two values of $\tilde{h}^*$, namely 0.19 (circles) and 0.25 (squares). Filled symbols are results of simulations, while empty symbols represent extrapolated results.  Parameters that are common to all simulations are: $c/c^* = 1,\, z = 1,$\, $ N_{k} = 200$ and $W\!i = 2.6$. The value of $\bar{X}^*_\text{eq}$ used for the calculation of $E$ at the various values of $N_b$ are given in Table~\ref{ParaValue}. Values of $b$, $\chi(b)$, $h^*$, $z^*$, $\lambda_1^*$ and $\dot \epsilon^*$ used for each of the simulated values of $N_b = \{6, 8, 10, 12 \}$, are calculated as per the procedure described in section III of the main paper. Lines through the data at these values of $N_b$ indicate extrapolation to the limit $1/\sqrt{200}$. }
\label{ExtrapolationStretchPhase}
\end{figure}

As described in section~\ref{SFG}, the successive fine-graining technique maintains the key experimental variables constant at each level of fine-graining. For the experimental results displayed in Fig.~\ref{FullStretchRelaxation}, these are: $\{ c/c^* = 1,\, z = 1, \, N_{k} = 200, \,W\!i = 2.6\}$. Note that the choice $N_{k} = 200$ represents our knowledge of the contour length $L$, and the persistence length $\lambda_\text{p}$ of $\lambda$-phage DNA. For each choice of $N_b$, the parameters, $b$, $\chi(b)$, $h^*$, $z^*$, $\lambda_1^*$ and $\dot \epsilon^*$ that correspond to this set of experimental values can be calculated as described in section~\ref{SFG}. A representative set of values of these parameters for various values of $N_b$, obtained for the case $\tilde{h}^* = 0.19$, is displayed in Table~\ref{ParaValue}, along with the values of $\bar{X}^*_\text{eq}$ used for the calculation of $E$.

\begin{figure}[!ht]
\centering
\includegraphics[width=8.5cm,height=!]{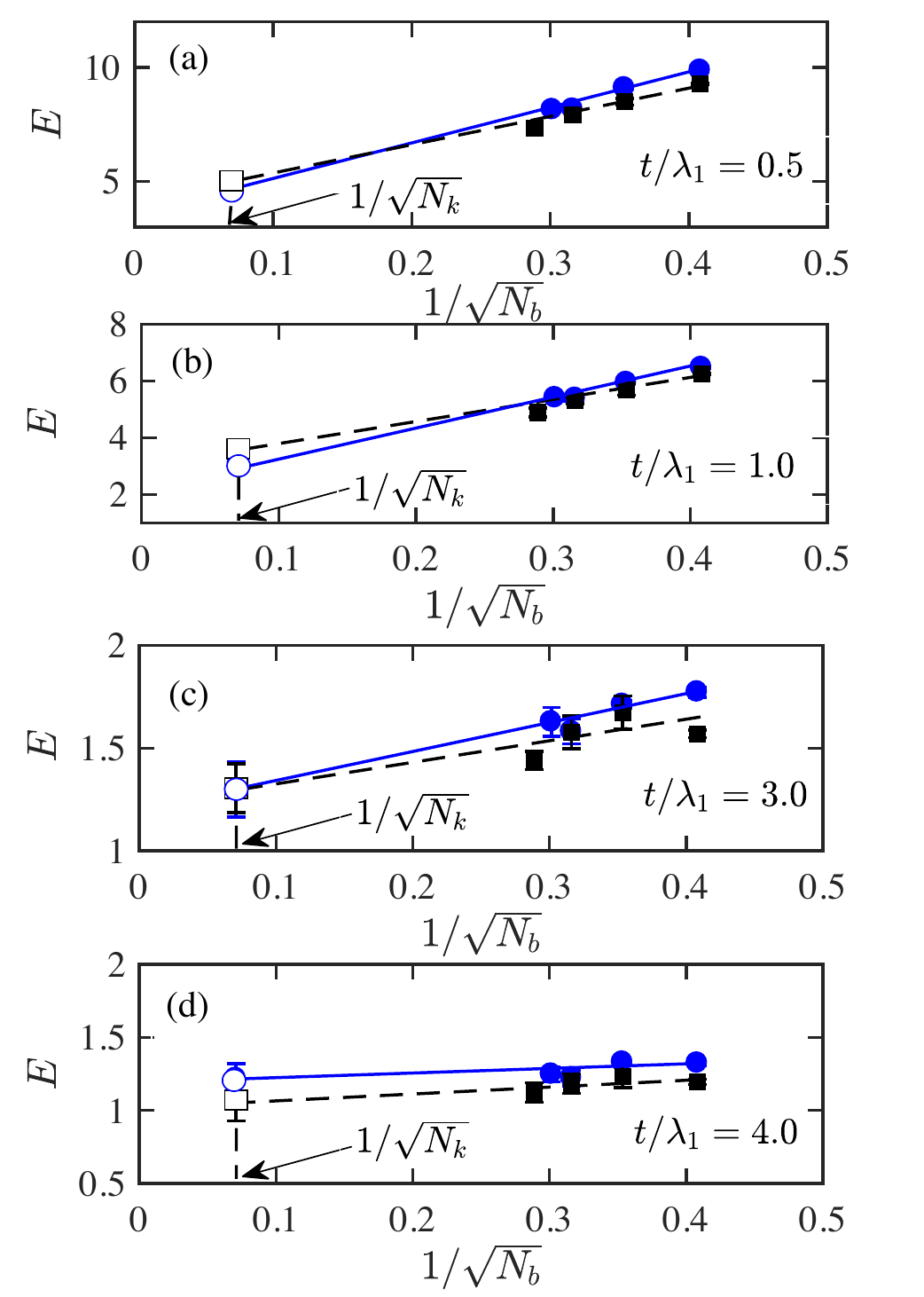}
\caption{ \footnotesize Illustration of the extrapolation procedure during the relaxation phase [(a) $t/\lambda_1 = 0.5$, (b) $t/\lambda_1 = 4.0$, (c) $t/\lambda_1 = 3.0$ and~(d) $t/\lambda_1 = 4.0$] for two values of $\tilde{h}^*$namely 0.19 (circles) and 0.25 (squares). Filled symbols are results of simulations, while empty symbols represent extrapolated results. Parameters that are common to all simulations are: $c/c^* = 1,\, z = 1,$\, $ N_{k} = 200$ and $W\!i = 2.6$. The value of $\bar{X}^*_\text{eq}$ used for the calculation of $E$ at the various values of $N_b$ are given in Table~\ref{ParaValue}. Values of $b$, $\chi(b)$, $h^*$, $z^*$, $\lambda_1^*$ and $\dot \epsilon^*$ used for each of the simulated values of $N_b = \{6, 8, 10, 12 \}$, are calculated as per the procedure described in section~III of the main paper. Lines through the data at these values of $N_b$ indicate extrapolation to the limit $1/\sqrt{200}$. }
\label{ExtrapolationRelaxationPhase}
\end{figure}

\begin{figure*}[t]
\centering
\includegraphics[width=17.8cm,height=!]{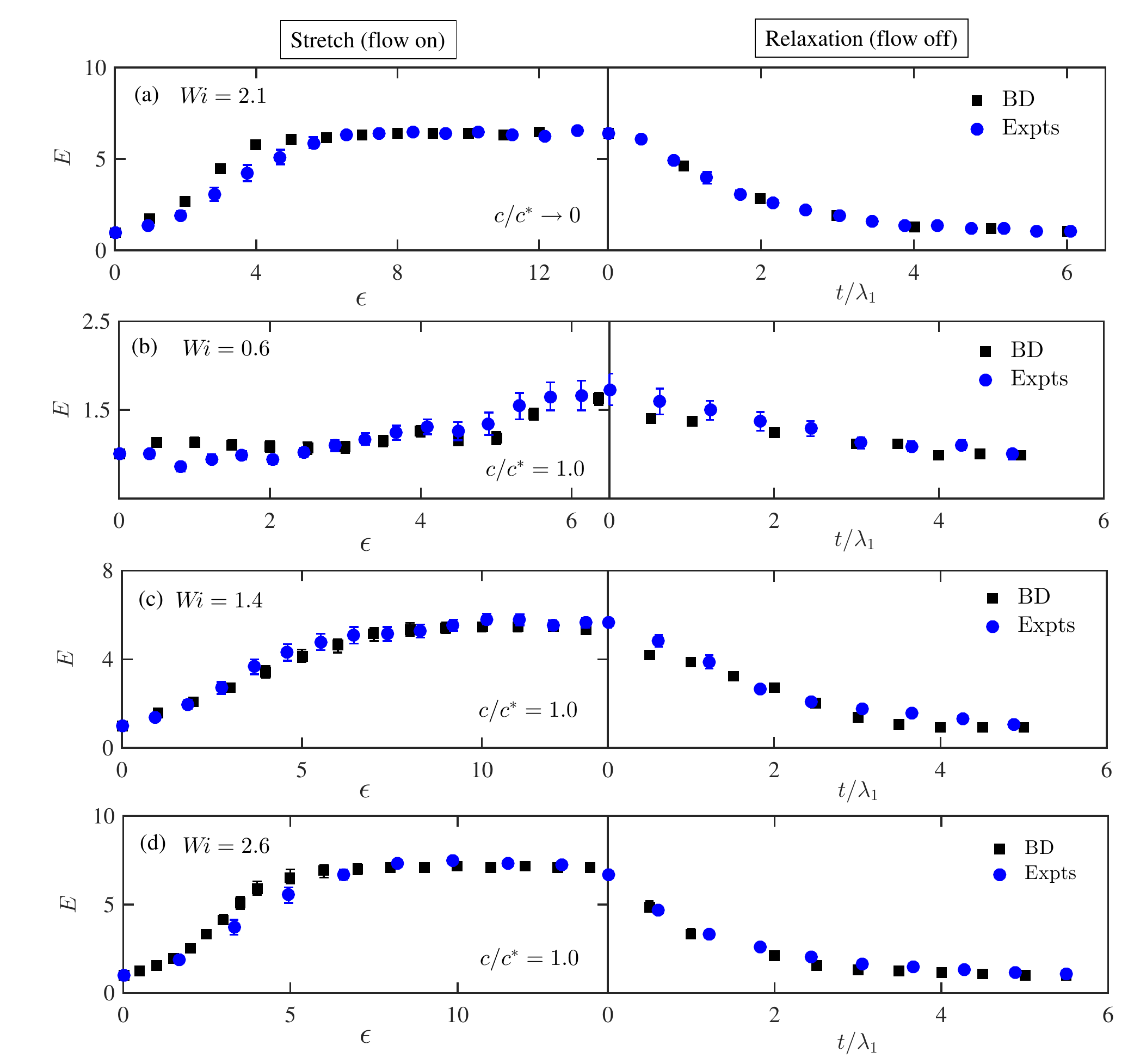}
\caption{ \footnotesize Comparison of the expansion factor $E= {\bar{X}} /{\bar{X}}_\text{eq}$ predicted by successive fine-graining with the experimental observations of Hsiao \etal~\cite{Kaiwen}. The top panel corresponds to a dilute solution at $W\!i = 2.1$. The remaining panels correspond to semidilute solutions at $c/c^* = 1$, and $W\!i = \{ 0.6, 1.4, 2.6 \}$, respectively. Simulations were carried out at fixed values of $z = 1$  and $N_{k} = 200$. Hsiao \etal~\cite{Kaiwen} have measured the values of $\bar{X}_\text{eq}$ and $\lambda_1$ at the start of each of their sets of experiments at the different values of Weissenberg number. They are used to plot the experimental data in the figure, and are reported here for convenience: [$W\!i, \bar{X}_\text{eq} (\mu\text{m}), \lambda_1$ (\text{s})]: [2.1 (dilute), $2.42 \pm 1.1$, $7.0$]; [0.6 (semidilute), $1.672 \pm 0.88$, $4.8$]; [1.4 (semidilute), $1.98 \pm 0.66$, $4.8$]; [2.6 (semidilute), $2.112 \pm 0.814$, $5.2$].}
\vskip-10pt
\label{StretchRelaxation}
\end{figure*}

Simulation predictions of the stretch ratio $E$ in a step strain followed by cessation of flow simulation, both in the stretch phase (at $\epsilon = 1.0$, \, $\epsilon = 4.0$, \,$\epsilon = 7.0$, and $\epsilon = 13.0$), and in the relaxation phase (at $t/\lambda_1 = 0.5$, \, $t/\lambda_1 = 1.0$, \,$t/\lambda_1 = 3.0$, and $t/\lambda_1 = 4.0$), at two different values of $\tilde{h}^*$, for a set of coarse-grained chains with $N_b = \{6, 8, 10, 12 \}$, are shown in Figs.~\ref{ExtrapolationStretchPhase} and~\ref{ExtrapolationRelaxationPhase}, respectively. In each case, data accumulated for these values of $N_b$ is extrapolated to the limit $(1/\sqrt{N_\text{k}}) = 1/\sqrt{200}$. Clearly, in all cases, the extrapolated value of the expansion factor $E$ is independent of the choice of value for $\tilde{h}^*$, within simulation error bars. As mentioned earlier, for the results to be truly parameter free, it is necessary to demonstrate independence of the extrapolated results from the choice of the constant $K$ in the narrow Gaussian potential  as well. In the supplementary material, we show that data accumulated for various values of $N_b$, at $W\!i = 2.6$ for two different values of $K$, extrapolate to a common value (within error bars) in the limit $(1/\sqrt{N_\text{k}})$. This implies that at $W\!i = 2.6$, in the stretch and relaxation phases,  local details of the chain (such as the nondimensional bead radius and the range of the excluded volume potential) are masked from the flow, even though the polymer chains are exposed to a flow field, and universal predictions independent of choice of parameter values are obtained. 

We can anticipate that at higher Weissenberg numbers, and large values of strain, as the flow penetrates down to the shortest length scales of the chains, the different values chosen for $\tilde{h}^*$ may get ``revealed", leading to predictions that are no longer parameter free. In the next section, we develop a simple scaling argument to obtain an estimate of the Weissenberg number at which this might happen. For all the values of $W\!i$, $\epsilon$ and $t/\lambda_1$ considered in the experiments of~\citet{Kaiwen}, however, we obtain parameter free predictions from the successive fine-graining procedure. 

\citet{Kaiwen} have carried out step strain followed by cessation of flow experiments, for an ultra-dilute solution ($c/c^* = 10^{-5}$) and for a semidilute solution ($c/c^* = 1$), for a range of different Weissenberg numbers. Predictions of the transient stretch ratio, obtained by carrying out the successive fine-graining procedure for a dilute solution with $c/c^* = 6.25 \times10^{-12}$ at $W\!i = 2.1$, and for a semidilute solution with $c/c^* = 1$ at $W\!i = \{ 0.6, 1.4, 2.6 \}$, at each of the measured values of $\epsilon$ in the stretch phase, and $t/\lambda_1$ in the relaxation phase, are shown in Fig.~\ref{StretchRelaxation}, and compared with the measurements of \citet{Kaiwen}. Clearly, the agreement between simulations and experiments is remarkable, and shows the usefulness of the successive fine-graining procedure in obtaining parameter free predictions that are in quantitative agreement with measurements. Further, they suggest that  coarse-grained Brownian dynamics simulations appear to be capable of capturing the important physics that determine the dynamics of semidilute solutions. 

\begin{figure}[!ht]
\centering 
\includegraphics[width=8.5cm,height=!]{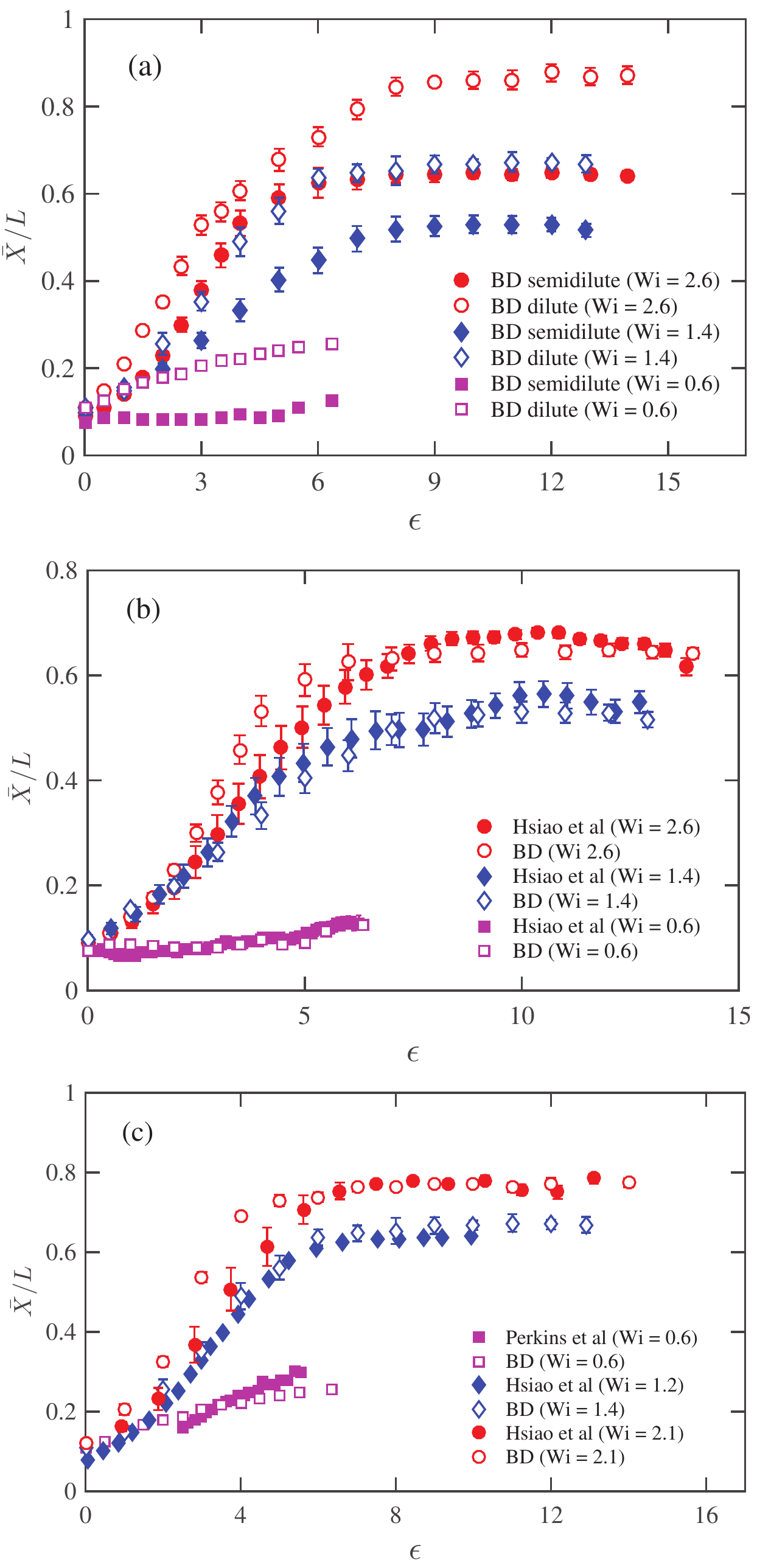}
\caption{\label{inhibstretch} \footnotesize  Transient polymer stretch in dilute and semidilute solutions at various values of the Weissenberg number. (a) Comparison of transient fractional extension $(\bar X/L)$ in planar extensional flow for dilute and semidilute solutions (at $c/c^* =1$) predicted by successive fine-graining. (b) Comparison of $(\bar X/L)$ for semidilute solutions predicted by successive fine-graining with experimental observations of~\citet{Kaiwen}. (c) Comparison of $(\bar X/L)$ for dilute solutions predicted by successive fine-graining with experimental observations of~\citet{Kaiwen} and~\citet{Perkins}. Note that $L=22\,\mu$m has been used to normalise the experimental values of stretch.}
\label{ComDiluteAndSemi}
\end{figure}

An important experimental observation by~\citet{Kaiwen} is that the average transient fractional extension in start-up of planar extensional flow  in a semidilute solution is much smaller than in a dilute solution, suggesting that interactions with surrounding chains restrains the stretching of chains. The formation of transient structures due to intermolecular interactions has been proposed in earlier experiments on semidilute solutions in shear flow~\cite{Hur,BabcockPRL,Harasim,Huber}. Fig.~\ref{ComDiluteAndSemi}(a) compares the prediction by successive fine-graining of $(\bar X/L)$ versus $\epsilon$, for a dilute solution (at $c/c^* = 6.25 \times10^{-12}$) and a semidilute solution (at $c/c^* =1$), for three different values of the Weissenberg number. Clearly, $(\bar X/L)$  is smaller for semidilute solutions than for dilute solutions at all values of $W\!i$ and $\epsilon$, suggesting that BD  simulations also exhibit the strong inhibition of chain stretching in semidilute solutions observed in experiments. The precise nature of the intermolecular interactions that lead to this phenomenon will be investigated further in the future. Fig.~\ref{ComDiluteAndSemi}(b) compares the successive fine-graining predictions of the average transient fractional extension in semidilute solutions, with the experimental observations of~\citet{Kaiwen}. This comparison is identical to the one carried out for semidilute solutions in Fig.~\ref{StretchRelaxation}. However, it is restricted to the stretching dynamics, and is in terms of the ratio $(\bar X/L)$ rather than $E$. Fig.~\ref{ComDiluteAndSemi}(c) compares the successive fine-graining predictions of $(\bar X/L)$ for dilute solutions with experimental observations. At  $W\!i = 0.6$, comparison is made with the measurements of~\citet{Perkins}. The comparison with the dilute solution measurements of~\citet{Kaiwen} for $W\!i = 2.1$ is identical to the comparison of stretching dynamics in Fig.~\ref{StretchRelaxation}, but is reported in terms of $(\bar X/L)$ rather than $E$. We have not carried out simulations at $W\!i = 1.2$, for which~\citet{Kaiwen} have reported experimental measurements. However, as seen in the figure, successive fine-graining predictions at $W\!i = 1.4$ are very close to the experimental values at $W\!i = 1.2$. Figs.~\ref{ComDiluteAndSemi}(b) and ~\ref{ComDiluteAndSemi}(c) once again reflect the quantitative accuracy with which successive fine-graining can predict transient chain stretch in extensional flows. 

\section{\label{blobology} Breakdown of successive fine graining}
It is possible to use scaling arguments based on blob theory to understand the observed independence from the choice of $\tilde{h}^*$, and to get an estimate of the value of $W\!i$ at which the successive fine graining scheme may be expected to breakdown. We first make a simple qualitative argument for a dilute solution in the good solvent limit $z \to \infty$ (where the thermal blob length scale is  expected to be of the order of monomer size) followed by a more general and detailed scaling analysis for both dilute and semidilute solutions below. 

In a dilute solution in the good solvent limit, a chain obeys self avoiding walk statistics, with the Flory exponent $\nu$, on all length scales. Consider the conformation of such a chain when the solution is subjected to extensional flow at a particular value of $W\!i$. At high values of strain $\epsilon$, the chain breaks up into a sequence of Pincus blobs of size $\xi_\text{P}$, which is the length scale at which the stretching energy in a chain segment becomes of order $k_{B}T$. Under these conditions, the conformation of the chain will be  rodlike on length scales above $\xi_\text{P}$, but for smaller length scales, chain segments will have equilibrium conformations. In other words, the flow does not ``penetrate'' the chain on length scales below $\xi_\text{P}$, and equilibrium conditions apply on these short length scales. The friction experienced by the chain as a whole is equal to that experienced by a blob-pole, and the friction coefficient of individual {monomers} is not relevant, since they are buried inside the blobs. Provided the conformations of chains with different local properties are the same on length scales large compared to the Pincus blob, their long time and large scale behavior will be identical. This is the reason why simulation results become independent of $\tilde{h}^*$ when sufficient degrees of freedom are taken into account. The transition length scale $\xi_\text{P}$ at which a chain switches from its equilibrium conformation to a deformed conformation depends on the Weissenberg number $W\!i$. 

An estimate of $\xi_\text{P}$ and the critical Weissenberg number $W\!i_\text{c}$ at which the successive fine graining technique can be expected to break down, under very general conditions, is obtained here in two parts. We first consider the case of dilute solutions, which helps to introduce the notation and establish the basic procedure for determining these quantities. A brief consideration of the key issues that are relevant in the case of semidilute solutions is given in this section, and because of the many regimes involved in this case, the details of the derivation of the various scaling laws are provided in appendix~\ref{sec:semidil}, and only the main results summarised in Table~\ref{tableblob}.

\subsection{\label{dilsol} Dilute solutions}

At equilibrium, the conformation of an isolated chain in a dilute solution is expected to breakup into a sequence of thermal blobs of diameter $\xi_\text{T}$, which is the length scale at which the total pairwise excluded volume interaction energy of all the monomers within a blob is of order $k_{B}T$. The chain obeys random walk (RW) statistics below $\xi_\text{T}$, while the thermal blobs themselves obey self-avoiding walk (SAW) statistics on larger length scales. On the other hand, since hydrodynamic interactions are present on all length scales, the chain exhibits Zimm dynamics. Within the blob scaling picture, the solvent quality $z$ is given by~\cite{JainPRL} 
\begin{equation}
\label{zblob}
z  = \frac{R^{0,\theta}_\text{eq}}{\xi_\text{T}}
\end{equation}
where, $R^{0,\theta}_\text{eq} = b_\text{K} N_\text{K}^{{1}/{2}}$, is the mean size of the chain in a dilute solution under $\theta$ conditions, with $b_\text{K}$ being the length of a monomer. The solvent quality can be viewed as a measure of the number of thermal blobs on a chain, $\mathcal{N}_\text{T}$, since one can show, $\mathcal{N}_\text{T} =  z^2 $. 

In the presence of extensional flow, the chain conformation is a blob pole (i.e., an aligned sequence of Pincus blobs) on large length scales, while within a Pincus blob, the chain conformation remains at equilibrium. At low values of $W\!i$ (when the Pincus blob size is large), we expect that there will be many thermal blobs within a Pincus blob. At sufficiently high values of $W\!i$, however, as the stretching energy of the chain increases, the Pincus blob size is expected to shrink below that of a thermal blob. These two conditions, i.e, $\xi_\text{P} > \xi_\text{T}$, and $\xi_\text{P} < \xi_\text{T}$, lead to different scaling considerations, as detailed below.

\noindent {\bf(i)} $\bm{\xi_\text{P} > \xi_\text{T}}$: 

If there are $m_\text{T,P}$ thermal blobs in the Pincus blob, then since the thermal blobs exclude each other, $\xi_\text{P} = \xi_\text{T} \, m_\text{T,P}^{\nu}$, and the Zimm relaxation time is $\lambda_\text{P} = \lambda_\text{T} \, m_\text{T,P}^{3\nu}$, where, $\lambda_\text{T}$ is the relaxation time of a thermal blob. Clearly, the flow penetrates a Pincus blob when $\lambda_\text{P} = {\dot \epsilon}^{-1}$. As a result
\[  \left( \frac{\xi_\text{P} }{\xi_\text{T} } \right)^3 = \frac{\lambda_\text{P} }{\lambda_\text{T}} = \left( \lambda_\text{T} \, {\dot \epsilon} \right)^{-1} \]
The longest (Zimm) relaxation time of the chain is $\lambda_{1} = \lambda_\text{T}  \mathcal{N}_\text{T}^{3\nu}$, while the mean equilibrium size of the chain is given by $R_\text{eq} = \xi_\text{T}  \mathcal{N}_\text{T}^{\nu}$. This implies, 
\[\frac{\lambda_{1} }{\lambda_\text{T}}  = \left( \frac{R_\text{eq}}{\xi_\text{T} } \right)^3 \]
From the definition of the Weissenberg number
\[ {W\!i} = \lambda_{1} \dot \epsilon = \left( \lambda_\text{T}  \, {\dot \epsilon} \right)  \mathcal{N}_\text{T}^{3\nu} = 
\left( \frac{\xi_\text{T} }{\xi_\text{P} } \right)^3  \mathcal{N}_\text{T}^{3\nu} = \left( \frac{R_\text{eq}}{\xi_\text{P} } \right)^3 \]
The size of the Pincus blob is consequently given by
\begin{equation}
\label{dilWic1}
\xi_\text{P}  = R_\text{eq} \, {W\!i}^{-\frac{1}{3}}
\end{equation}
For $W\!i \sim \mathcal{O}(1)$, the entire chain is within a Pincus blob. At a critical value of the Weissenberg number, $W\!i_\text{c}$, the dimension of the Pincus blob would become of the order of the thermal blob size, i.e, $\xi_\text{P} = \xi_\text{T}$. Since $\mathcal{N}_\text{T} =  z^2 $ implies
\begin{equation}
\label{zblob1}
\frac{R_\text{eq}}{\xi_\text{T}} = z^{2\nu}
\end{equation}
It follows from Eq.~(\ref{dilWic1}) that
\begin{equation}
\label{Wic1}
W\!i_\text{c}  = z^{6\nu}
\end{equation}
As the thermal blobs get smaller with increasing solvent quality, it takes a higher value of the critical Weissenberg number before the Pincus blob penetrates the thermal blob. 

\noindent {\bf(ii)} $\bm{\xi_\text{P} < \xi_\text{T}}$: 

If there are $g_\text{P}$ monomers in a Pincus blob, then since RW statistics are obeyed within a thermal blob, $\xi_\text{P} = b_\text{K} \, g_\text{P}^{{1}/{2}}$, and the Zimm relaxation time of the Pincus blob is $\lambda_\text{P} = \lambda_{0} \, g_\text{P}^{{3}/{2}}$, where, $\lambda_{0}$ is the monomer relaxation time. Since the flow penetrates a Pincus blob when $\lambda_\text{P} = {\dot \epsilon}^{-1}$
\[  \left( \frac{\xi_\text{P} }{b_\text{K}} \right)^3 = \frac{\lambda_\text{P} }{\lambda_{0}} = \left( \lambda_{0} \, {\dot \epsilon} \right)^{-1} \]
If there are $g_\text{T} = (N_\text{K}/\mathcal{N}_\text{T})$ monomers in a thermal blob, then $\xi_\text{T} = b_\text{K} \, g_\text{T}^{{1}/{2}}$, and the Zimm relaxation time of a thermal blob is $\lambda_\text{T} = \lambda_{0} \, g_\text{T}^{{3}/{2}}$. It follows that
\[ \frac{\lambda_\text{T} }{\lambda_{0}} = \left( \frac{\xi_\text{T} }{b_\text{K}} \right)^3 \]
which implies
\[ \frac{\lambda_{1} }{\lambda_{0}} =  \frac{\lambda_{1} }{\lambda_\text{T}}  \frac{\lambda_\text{T} }{\lambda_{0}} = \left( \frac{R_\text{eq}}{\xi_\text{T} } \right)^3 \left( \frac{\xi_\text{T} }{b_\text{K}} \right)^3 = \left( \frac{R_\text{eq}}{b_\text{K}} \right)^3 \]
Since 
\begin{align} 
\label{req1}
R_\text{eq} & = \xi_\text{T}  \mathcal{N}_\text{T}^{\nu} =  b_\text{K} \, g_\text{T}^{{1}/{2}} \mathcal{N}_\text{T}^{\nu} \nonumber =  b_\text{K} N_\text{K}^{{1}/{2}}\mathcal{N}_\text{T}^{{(2\nu-1)}/{2}} \\ & =  b_\text{K} N_\text{K}^{{1}/{2}} z^{2\nu-1}
\end{align}
we get
\begin{equation*}
\resizebox{0.475\textwidth}{!}{$ 
{W\!i} = \lambda_{1} \dot \epsilon = \left( \lambda_{0} \, {\dot \epsilon} \right) \left( \frac{R_\text{eq}}{b_\text{K}} \right)^3 = 
\left( \frac{b_\text{K}}{\xi_\text{P} } \right)^3  \, \left( \frac{R_\text{eq}}{b_\text{K}} \right)^3= \left( \frac{R_\text{eq}}{\xi_\text{P} } \right)^3 
$}   
\end{equation*}
We see that in this case as well, the size of the Pincus blob is given by
\begin{equation}
\label{dilWic2}
\xi_\text{P}  = R_\text{eq} \, {W\!i}^{-\frac{1}{3}}
\end{equation}
For sufficiently large Weissenberg numbers, the dimension of the Pincus blob would become of the order of monomer size, at which point the local chain details would no longer be shielded from the flow. Thus, for $\xi_\text{P} = b_K$,  Eqs.~(\ref{req1}) and~(\ref{dilWic2}) imply that the successive fine graining procedure would breakdown at 
\begin{equation}
\label{dilWic3}
 W\!i_\text{c} = N_\text{K}^{{3}/{2}} z^{6\nu-3}
\end{equation}
Since $z \sim N_\text{K}^{{1}/{2}}$, it follows that
\[ W\!i_\text{c} \sim N_\text{K}^{3\nu} \] 
This is inline with our expectation that universal behaviour is exhibited until higher Weissenberg numbers for longer chains. For $\lambda$-phage DNA, with $N_\text{K} \approx 200$, this implies $W\!i_\text{c}  \sim \mathcal{O}(10^3)$ to $ \mathcal{O}(10^4)$. It must be borne in mind that this is a very rough estimate, based on scaling arguments, which do not predict pre-factors. It is also in some some sense an upper bound on the Weissenberg number, since the influence of the flow on the local details could occur when $\xi_\text{P}$ is of the order of many  monomer sizes. In the results of successive fine graining for dilute DNA solutions reported in Ref.~\cite{Sunthar}, it was observed that extrapolated results were parameter free at $W\!i =2$ for all strains, while universality broke down at $W\!i =55$, for high values of $\epsilon$. 

\begin{table*}[!ht]
\caption{\label{tableblob} Critical Weissenberg number at which the flow penetrates the Pincus blob. Various scaling regimes are determined by the relative magnitudes of the Pincus blob, the thermal blob and the correlation blob (for semidilute solutions). When the magnitude of the Pincus blob is equal to monomer size ($\xi_\text{P} = b_\text{K}$), the successive fine graining technique is expected to breakdown.}  
\vskip10pt
\centering       
\bgroup
\setlength{\tabcolsep}{0.5em}
{\def\arraystretch{1.7}
\begin{tabular}{| c | c | c | c | c |}    
\hline                        
  \multicolumn{5}{|c|}{Dilute}   \\
\hline \hline  
  & $W\!i$ &  $R_\text{eq}$ & Condition & $W\!i_\text{c} $ \\
 \hline    
$ \xi_\text{T} < \xi_\text{P}$ & $R_\text{eq}^3/\xi_\text{P}^3$ & $\xi_\text{T} \, z^{2\nu}$& $\xi_\text{P} = \xi_\text{T}$ & $z^{6\nu}$ \\
 \hline  
$ \xi_\text{P} < \xi_\text{T}$ & $R_\text{eq}^3/\xi_\text{P}^3$  &$ b_\text{K} N_\text{K}^{\frac{1}{2}} \, z^{2\nu-1}$ & $\xi_\text{P} = b_\text{K}$ & $ N_\text{K}^{3\nu}$ \\
 \hline  
\end{tabular}
\vskip20pt
\begin{tabular}{| c | c | c | c | c | c |}    
\hline 
  \multicolumn{6}{|c|}{Semidilute}   \\
\hline \hline  
&  & $W\!i$ &  $R_\text{eq}$ & Condition & $W\!i_\text{c} $ \\
 \hline    
 \multirow{2}{*}{$ \xi_\text{T} < \xi_\text{c}$} & $ \xi_\text{T} <  \xi_\text{c} < \xi_\text{P}$ & $R_\text{eq}^4/\xi_\text{P}^4$ & $\xi_\text{c} \left( c / c^* \right)^{\tfrac{1}{6 \nu - 2}}$& $\xi_\text{P} = \xi_\text{c}$ & $ \left( c / c^* \right)^{\tfrac{2}{3 \nu - 1}}$ \\
\cline{2-6} 
\multirow{2}{*}{ $c^{*} < c < c^{**} $}  & $ \xi_\text{T} <  \xi_\text{P} < \xi_\text{c}$ & $\left( R_\text{eq}/\xi_\text{P}\right)^3 \left( c / c^* \right)^{\tfrac{1}{6 \nu - 2}}$ & $\xi_\text{T} \, z^{2\nu} \left( c / c^* \right)^{- \tfrac{2\nu-1}{6 \nu - 2}}$& $\xi_\text{P} = \xi_\text{T}$ & $z^{6\nu} \left( c / c^* \right)^{- \tfrac{3\nu-2}{3 \nu - 1}} $ \\
\cline{2-6}  
 & $ \xi_\text{P} < \xi_\text{T} < \xi_\text{c}$ & $\left( R_\text{eq}/\xi_\text{P}\right)^3 \left( c / c^* \right)^{\tfrac{1}{6 \nu - 2}}$ & $b_\text{K} N_\text{K}^{\frac{1}{2}} \,  z^{2\nu-1} \left( c / c^* \right)^{- \tfrac{2\nu-1}{6 \nu - 2}}$& $\xi_\text{P} = b_\text{K}$ & $N_\text{K}^{3\nu} \left( c / c^* \right)^{- \tfrac{3\nu-2}{3 \nu - 1}} $ \\
 \hline  
  \hline 
 \multirow{2}{*}{$ \xi_\text{c} < \xi_\text{T}$} & $ \xi_\text{c} <  \xi_\text{T} < \xi_\text{P}$ & $R_\text{eq}^4/\xi_\text{P}^4$ & $\xi_\text{T} \, z$& $\xi_\text{P} = \xi_\text{T}$ & $ z^{4}$ \\
\cline{2-6} 
\multirow{2}{*}{ $c^{**} < c  $}  & $ \xi_\text{c} <  \xi_\text{P} < \xi_\text{T}$ & $R_\text{eq}^4/\xi_\text{P}^4$ & $\xi_\text{c} \left( c / c^* \right)$& $\xi_\text{P} = \xi_\text{c}$ & $\left( c / c^* \right)^{4} $ \\
\cline{2-6}  
 & $ \xi_\text{P} < \xi_\text{c} < \xi_\text{T}$ & $\left( R_\text{eq}/\xi_\text{P}\right)^3 \left( c / c^* \right)$ & $b_\text{K} N_\text{K}^{\frac{1}{2}} $& $\xi_\text{P} = b_\text{K}$ & $N_\text{K}^{\frac{3}{2}} \left( c / c^* \right) $ \\
 \hline  
\end{tabular}
}
\egroup
\end{table*}

\subsection{\label{semidilsol} Semidilute solutions}
At equilibrium, the onset of the semidilute regime occurs at the concentration $c^*$, where chains just begin to overlap each other. Within the blob ansatz, at higher concentrations, chain conformations breakup into a sequence of correlation blobs of diameter $\xi_\text{c}$, with sections of chains within a blob behaving as they would in a dilute solution. The correlation blobs themselves are assumed to be space filling, so the solution behaves like a melt of correlation blobs on length scales larger than  $\xi_\text{c}$. Since dilute solution dynamics are observed within a correlation blob, chain segments within these blobs are further subdivided into thermal blobs, whose magnitude and number depend on the quality of the solvent. On  length scales above $\xi_\text{c}$, since melt dynamics are observed, chains obey random walk statistics and Rouse dynamics~\cite{Rubinstein}. A phase diagram in the $\{z, c/c^*\}$ space, with a derivation of the various scaling laws that operate in the different regimes, has been presented recently in Ref.~\cite{JainPRL}.

For concentrations less than or equal to $c^*$, since the entire chain is within a correlation blob, the same arguments as those used for dilute solutions above would apply at the onset of flow. For $c/c^* > 1$, however, we expect that there will be a subtle interplay between the different blob length scales that are present, with different scaling laws governing the different regimes. At equilibrium, one can distinguish two cases: (i) $\xi_\text{T} < \xi_\text{c}$, which would hold for $c^* < c < c^{**}$, and (ii) $\xi_\text{c} < \xi_\text{T}$, which would hold for $c^{**} < c$ (note that $c^{**}$ is defined as the concentration at which $\xi_\text{c} = \xi_\text{T}$). Once extensional flow is switched on, the magnitude of $\xi_\text{P}$ relative to $\xi_\text{T}$ and  $\xi_\text{c}$ depends on the value of $W\!i$, and this in turn determines which microscopic physics is relevant. 

There are three possible scenarios. Consider the case, $\xi_\text{T} < \xi_\text{c}$. At low extension rates, there will be many correlation blobs within a Pincus blob with their number decreasing as the Pincus blobs decrease in size with increasing strain rate, until the size of the Pincus blob becomes of the order of the correlation blob size. At higher Weissenberg numbers, the Pincus blobs become smaller than the correlation blobs, until they become of order of the thermal blob size. Eventually, at sufficiently high Weissenberg numbers, the Pincus blob penetrates the thermal blob, and its size becomes comparable to the monomer size. We anticipate that the successive fine graining procedure will breakdown at this point, since the local details of the chain would be exposed to the flow. For the case, $\xi_\text{c} < \xi_\text{T}$, the roles of the correlation and thermal blobs are interchanged in the above sequence of events. The critical Weissenberg numbers at which the Pincus blob size becomes equal to the correlation and thermal blob sizes and to the monomer size can be estimated in  the various cases, as shown in 
appendix~\ref{sec:semidil}. 

All the scaling expressions derived here for dilute and semidilute solutions  are summarised in Table~\ref{tableblob}. As mentioned earlier, pre-factors cannot be determined within the framework of scaling arguments, but must rather be determined by careful simulations that explore the threshold Weissenberg number at which results are no longer parameter free. 

\section{\label{conclu}Conclusions}

The dynamics of DNA molecules in semidilute solutions undergoing planar extensional flow has been simulated using a coarse-grained bead-spring chain model which incorporates hydrodynamic and excluded volume interactions. When applied to semidilute solutions, the successive fine-graining methodology is shown to lead to parameter-free predictions for a range of Weissenberg numbers and Hencky strain units, as was observed previously for dilute solutions~\cite{Sunthar,PrabhakarSFG,saadat2015molecular}. A systematic comparison of simulation predictions with the experimental observations of~\citet{Kaiwen}, of the response of individual chains to step-strain deformation followed by cessation of flow, shows that the successive fine graining technique gives quantitatively accurate predictions in the experimentally explored range of Weissenberg numbers. In agreement with experimental observations, simulations indicate that the transient chain stretch following a step strain deformation is much smaller in semidilute solutions than in dilute solutions. 

The current work has been focussed on comparing simulation predictions with the experimental observations of \citet{Kaiwen}, which have all been carried out at $c/c^*=1$. Clearly, a thorough examination of the influence of concentration on the stretching and relaxation dynamics, particularly with a view to understanding the nature of the interchain interactions that lead to restriction in chain stretching,  is required in the future. 

The simple scaling analysis based on the blob picture in section~\ref{blobology} suggests that the relative magnitudes of Pincus and correlation blobs depend on the key variables that determine semidilute solution dynamics: $\{R_\text{eq}, L, c/c^*, W\!i \}$. The interplay between these two length scales in turn influences the manner in which hydrodynamic interactions are screened, which is at the heart of the rich physics observed in semidilute polymer solutions. By making it possible to study long chain behaviour by simulating shorter chains, the method of successive fine graining provides a means of studying local chain structure as a function of these variables (via, for instance, the dynamic structure factor).  Future studies in this direction would give insight into their influence on the screening of hydrodynamic interactions.
\vskip10pt
\noindent \textbf{SUPLEMENTARY MATERIAL}
\vskip10pt
\noindent See supplementary material at [URL to be inserted by AIP] for details of the following: integration scheme for the stochastic differential equation; simulation procedure and protocols; particular forms of the spring force, and the hydrodynamic interaction tensor used here; procedure for determining the longest relaxation time; comparison of the results of successive fine graining for $z=0.7$ and $z=1.0$; and demonstration of the independence from choice of value for $K$.

\begin{acknowledgments}
This research was supported under the Australian Research Council's Discovery Projects funding scheme (project DP120101322). It was undertaken with the assistance of resources provided at the NCI National Facility systems at the Australian National University through the National Computational Merit Allocation Scheme supported by the Australian Government, and was supported by a Victorian Life Sciences Computation Initiative (VLSCI) Grant number VR0010 on its Peak Computing Facility at the University of Melbourne, an initiative of the Victorian Government, Australia.   
\end{acknowledgments}

\begin{appendix}

\section{\label{sec:semidil} Scaling regimes for $W\!i_\text{c} $ in semidilute solutions}

\noindent {\bf(i)} $\bm{\xi_\text{T} < \xi_\text{c}}$ (RW statistics below $\xi_\text{T}$ and above $\xi_\text{c}$; SAW statistics above $\xi_\text{T}$ and below $\xi_\text{c}$; Zimm dynamics below $\xi_\text{c}$; Rouse dynamics above $\xi_\text{c}$).
\vskip10pt
\noindent (a) \underline{$\xi_\text{T} < \xi_\text{c} < \xi_\text{P}$}
\vskip10pt
At sufficiently low Weissenberg numbers, if there are $m_\text{c,P}$ correlation blobs in a Pincus blob, then $\xi_\text{P} = \xi_\text{c} \, m_\text{c,P}^{1/2}$, and the Rouse relaxation time is $\lambda_\text{P} = \lambda_\text{c} \, m_\text{c,P}^{2}$, where $\lambda_\text{c}$ is the Zimm relaxation time of a correlation blob. Since the flow penetrates a Pincus blob when $\lambda_\text{P} = {\dot \epsilon}^{-1}$
\[  \left( \frac{\xi_\text{P} }{ \xi_\text{c}} \right)^4 = \frac{\lambda_\text{P} }{ \lambda_\text{c}} = \left( \lambda_\text{c} \, {\dot \epsilon} \right)^{-1} \]
In the semidilute regime, the longest (Rouse) relaxation time of the chain is $\lambda_{1} = \lambda_\text{c} \, \mathcal{N}_\text{c}^{2}$, where $\mathcal{N}_\text{c}$ is the number of correlation blobs in a chain, and the mean equilibrium size of the chain is given by $R_\text{eq} =  \xi_\text{c} \, \mathcal{N}_\text{c}^{1/2}$. It follows that
\[ {W\!i} = \lambda_{1} \dot \epsilon = \left( \lambda_\text{c} \, {\dot \epsilon} \right) \, \mathcal{N}_\text{c}^{2} = 
\left( \frac{\xi_\text{c} }{\xi_\text{P} } \right)^4  \,  \mathcal{N}_\text{c}^{2} = \left( \frac{R_\text{eq}}{\xi_\text{P} } \right)^4 \]
At low Weissenberg numbers, the size of the Pincus blob is consequently given by
\begin{equation}
\label{semdilWic1}
\xi_\text{P}  = R_\text{eq} \, {W\!i}^{-\frac{1}{4}} 
\end{equation} 
The Pincus blob in a semidilute solution (in this sub-case), appears to decrease more slowly in size than in a dilute solution. The number of correlation blobs in a chain can be related to the scaled concentration through~\cite{JainPRL} 
\begin{equation}
\label{nc}
\mathcal{N}_\text{c} =   \left( \frac{c}{c^*} \right)^{\tfrac{1}{3 \nu - 1}}  
\end{equation}
As a result, $R_\text{eq} =  \xi_\text{c} \left( {c}/{c^*} \right)^{{1}/({6 \nu - 2})} $, and the Weissenberg number at which $\xi_\text{P} = \xi_\text{c}$ is given by
\begin{equation}
W\!i_\text{c} = \left( \frac{c}{c^*} \right)^{ \tfrac{2}{3 \nu - 1}} 
\end{equation}
At this Weissenberg number the conformation of a typical chain is a blob pole, with the blobs representing both the length scale at which the stretching energy is of order $k_{B}T$ (Pincus blob), and the length scale at which hydrodynamic and excluded volume interactions are screened (correlation blob). Pan et al.~\cite{pan2014universal} have shown that for DNA solutions, the unentangled semidilute regime appears to extend to roughly $c / c^* = 5$. The range of Weissenberg numbers at which the two blob sizes become equal is then (for $\nu =0.6$), $W\!i \sim \mathcal{O}(1)$ to $ \mathcal{O}(55)$ for $1 \le \left( c / c^* \right) \le 5$. 
\vskip10pt
\noindent (b) \underline{$\xi_\text{T} < \xi_\text{P} < \xi_\text{c}$}
\vskip10pt
At higher Weissenberg numbers, there will be several Pincus blobs within a correlation blob. Since the Pincus blobs within a correlation blob are expected to form a blob pole, we anticipate the correlation blobs to be anisotropic in structure, with width $\sim \xi_\text{P}$, but length of order several $\xi_\text{P}$. A careful examination of the different blob length scales that are present in a semidilute solution subjected to extensional flow, and the resultant chain conformations, has been carried out  recently by Prabhakar et al.~\cite{Prabhakar16}. However, we are interested in the equilibrium conditions that exist within a Pincus blob, and in equilibrium chain/blob dimensions and relaxation times. 

If there are $m_\text{T,P}$ thermal blobs in the Pincus blob, then $\xi_\text{P} = \xi_\text{T} \, m_\text{T,P}^{\nu}$, and the Zimm relaxation time is $\lambda_\text{P} = \lambda_\text{T} \, m_\text{T,P}^{3\nu}$. Since the flow penetrates a Pincus blob when $\lambda_\text{P} = {\dot \epsilon}^{-1}$
\[  \left( \frac{\xi_\text{P} }{\xi_\text{T} } \right)^3 = \frac{\lambda_\text{P} }{\lambda_\text{T}} = \left( \lambda_\text{T} \, {\dot \epsilon} \right)^{-1} \]
Using similar arguments to those above, one can show that
\[  \frac{\lambda_\text{c} }{\lambda_\text{T}} = \left( \frac{\xi_\text{c} }{\xi_\text{T} } \right)^3 \]
As a result,
%
\begin{align*}  
{W\!i} & = \lambda_{1} \dot \epsilon = \lambda_\text{c} \mathcal{N}_\text{c}^{2} \, {\dot \epsilon} = \left( \lambda_\text{T}  \, {\dot \epsilon} \right) \left( \frac{\lambda_\text{c}}{\lambda_\text{T}} \right)  \mathcal{N}_\text{c}^{2} \nonumber \\ & = \left( \frac{\xi_\text{T} }{\xi_\text{P} } \right)^3  \left( \frac{\xi_\text{c}}{\xi_\text{T}} \right)^3 \mathcal{N}_\text{c}^{2} = \left( \frac{\xi_\text{c}}{\xi_\text{P} } \right)^3  \mathcal{N}_\text{c}^{2}
\end{align*} 
%
Since $ \xi_\text{c} = R_\text{eq} \, \mathcal{N}_\text{c}^{\, -\frac{1}{2}}$, one can show that 
\begin{equation}
\label{xicnc}
\xi_\text{c}^3 \mathcal{N}_\text{c}^2 = R_\text{eq}^3 \left( \frac{c}{c^*} \right)^{ \tfrac{1}{2(3 \nu - 1)}}  
\end{equation}
where Eqn.~(\ref{nc}) has been used. The size of the Pincus blob is consequently given by
\begin{equation}
\label{semdilWic2}
\xi_\text{P}  = {W\!i}^{-\frac{1}{3}} \,  R_\text{eq} \left( \frac{c}{c^*} \right)^{ \tfrac{1}{6(3 \nu - 1)}} 
\end{equation}
We are interested in determining the Weissenberg number at which $\xi_\text{P} = \xi_\text{T}$. In order to do so, it is necessary to relate $R_\text{eq}$ to $\xi_\text{T}$. If there are $m_\text{T,c}$ thermal blobs in a correlation blob, then $R_\text{eq} = \xi_\text{c} \mathcal{N}_\text{c}^{\, \frac{1}{2}} = \xi_\text{T} m_\text{T,c}^{\nu} \, \mathcal{N}_\text{c}^{\, \frac{1}{2}}$. In the double crossover region, Jain et al.~\cite{JainPRL} have derived the following expressions for $m_\text{T,c}$ as a function of the solvent quality and the scaled concentration
\[ m_\text{T,c} = z^2  \left( \frac{c}{c^*} \right)^{- \tfrac{1}{3 \nu - 1}} \]
It follows that,
\begin{equation}
\label{semdilreq1}
R_\text{eq} = \xi_\text{T}  z^{2\nu} \left( \frac{c}{c^*} \right)^{- \tfrac{2\nu-1}{6 \nu - 2}} 
\end{equation}
and the critical Weissenberg number at which the Pincus blob and the thermal blob length scales are identical is given by
\begin{equation}
W\!i_\text{c} = z^{6\nu} \left( \frac{c}{c^*} \right)^{ - \tfrac{3\nu - 2}{3 \nu - 1}} 
\end{equation}
This reduces to the expression for the critical Weissenberg number for $\xi_\text{P} = \xi_\text{T}$ in dilute solutions (Eqn.~(\ref{Wic1})), when $(c/c^*) =1$. 

An estimate of the critical Weissenberg number can be obtained from the following arguments. If we assume that the unentangled semidilute regime for DNA is in the range~\cite{pan2014universal} $1< (c / c^*) < 5$, then the number of correlation blobs in this range (for $\nu =0.6$) is  $1< \mathcal{N}_\text{c} < 8$ (from Eqn.~(\ref{nc})). Since the number of thermal blobs must be greater than the number of correlation blobs in this sub-case, we assume that $\mathcal{N}_\text{T} = 16$ (at least two thermal blobs in each correlation blob). This implies $z=4$ (since $z=\sqrt{\mathcal{N}_\text{T}}$). Substituting these numbers into Eqn.~(\ref{nc}) leads to $W\!i_\text{c} \sim \mathcal{O}(1)$ to $ \mathcal{O}(220)$, for $1 \le \left( c / c^* \right) \le 5$. 
\vskip10pt
\noindent (c) \underline{$\xi_\text{P} < \xi_\text{T} < \xi_\text{c}$}
\vskip10pt
If there are $g_\text{P}$ monomers in a Pincus blob, then, $\xi_\text{P} = b_\text{K} \, g_\text{P}^{{1}/{2}}$, and the Zimm relaxation time of the Pincus blob is $\lambda_\text{P} = \lambda_{0} \, g_\text{P}^{{3}/{2}}$. Since the flow penetrates a Pincus blob when $\lambda_\text{P} = {\dot \epsilon}^{-1}$
\[  \left( \frac{\xi_\text{P} }{b_\text{K}} \right)^3 = \frac{\lambda_\text{P} }{\lambda_{0}} = \left( \lambda_{0} \, {\dot \epsilon} \right)^{-1} \]
Clearly
\[  \frac{\lambda_1}{\lambda_\text{0}} = \frac{\lambda_1}{\lambda_\text{c}} \frac{\lambda_\text{c}}{\lambda_\text{T}} \frac{\lambda_\text{T}}{\lambda_\text{0}} = \mathcal{N}_\text{c}^{2} \left( \frac{\xi_\text{c} }{\xi_\text{T} } \right)^3  \left( \frac{\xi_\text{T}}{b_\text{K}} \right)^3 = \left( \frac{\xi_\text{c}}{b_\text{K} } \right)^3  \mathcal{N}_\text{c}^{2} \]
It follows that
\begin{align*} 
{W\!i} & = \lambda_{1} \dot \epsilon = (\lambda_\text{0} {\dot \epsilon})  \left( \frac{\lambda_1}{\lambda_\text{0}} \right) \nonumber \\ &=  \left( \frac{b_\text{K}}{\xi_\text{P} } \right)^3 \left( \frac{\xi_\text{c}}{b_\text{K}} \right)^3 \mathcal{N}_\text{c}^{2} 
= \left( \frac{\xi_\text{c}}{\xi_\text{P} } \right)^3  \mathcal{N}_\text{c}^{2}
\end{align*}
Using Eqn.~(\ref{xicnc}), we can find the dependence of the size of the Pincus blob on the Weissenberg number to be
\begin{equation}
\label{xipsemidil1}
\xi_\text{P}  = {W\!i}^{-\frac{1}{3}} \,  R_\text{eq} \left( \frac{c}{c^*} \right)^{ \tfrac{1}{6(3 \nu - 1)}} 
\end{equation}
From Eqn.~(\ref{zblob}), we see that $\xi_\text{T} = b_\text{K} N_\text{K}^{1/2} \, z^{-1}$. Combined with Eqn.~(\ref{semdilreq1}) for $R_\text{eq}$, this leads to
\begin{equation}
\label{semdilreq2}
R_\text{eq} = b_\text{K}  N_\text{K}^{1/2}  z^{2\nu-1} \left( \frac{c}{c^*} \right)^{- \tfrac{2\nu-1}{6 \nu - 2}} 
\end{equation}
Substituting Eqn.~(\ref{semdilreq2}) into Eqn.~(\ref{xipsemidil1}), and setting $\xi_\text{P} = b_\text{K}$, we find the critical Weissenberg number at which the successive fine graining method is expected to break down to be
\begin{equation}
\label{semidilWic3}
 W\!i_\text{c} = N_\text{K}^{{3}/{2}} z^{6\nu-3} \left( \frac{c}{c^*} \right)^{ - \tfrac{3\nu - 2}{3 \nu - 1}} 
\end{equation}
This reduces to the expression for $W\!i_\text{c}$ for dilute solutions (Eqn.~(\ref{dilWic3})), when $(c/c^*) =1$. Since $z \sim N_\text{K}^{{1}/{2}}$, it follows that
\[ W\!i_\text{c} \sim N_\text{K}^{3\nu} \left( \frac{c}{c^*} \right)^{ - \tfrac{3\nu - 2}{3 \nu - 1}}  \] 
Assuming $N_\text{K} = 200$ for DNA, this leads to (for $\nu =0.6$) $W\!i_\text{c} \sim \mathcal{O}(10^3)$ to $ \mathcal{O}(10^4)$ for $1 \le \left( c / c^* \right) \le 5$. 

\vskip10pt
\noindent {\bf(ii)} $\bm{\xi_\text{c} < \xi_\text{T}}$ (RW statistics on all length scales; Zimm dynamics below $\xi_\text{c}$; Rouse dynamics above $\xi_\text{c}$). 
\vskip10pt
We consider only solutions where $N_\text{b}$ and $c$ are not large enough for entanglements to play a role, and that further crossover to reptation dynamics does not need to be taken into account. It should be pointed out that~\citet{Kaiwen} have not carried out any experiments in this regime, nor have we carried out any simulations. Nevertheless, the results are presented here for the sake of completeness; basically as a tabulation of critical Weissenberg numbers  at which local details would begin to effect predictions of coarse-grained models (which is essentially what is implied by the breakdown of successive fine graining). The three scenarios in this case are discussed in turn below. 
\vskip10pt
\noindent (a) \underline{$\xi_\text{c} < \xi_\text{T} < \xi_\text{P}$}
\vskip10pt
If $m_\text{c,P}$ is the number of correlation blobs in a Pincus blob, then $\xi_\text{P} = \xi_\text{c} \, m_\text{c,P}^{1/2}$, and the Rouse relaxation time is $\lambda_\text{P} = \lambda_\text{c} \, m_\text{c,P}^{2}$. Since the flow penetrates a Pincus blob when $\lambda_\text{P} = {\dot \epsilon}^{-1}$,
\[  \left( \frac{\xi_\text{P} }{ \xi_\text{c}} \right)^4 = \frac{\lambda_\text{P} }{ \lambda_\text{c}} = \left( \lambda_\text{c} \, {\dot \epsilon} \right)^{-1} \]
From the scaling expressions for the Rouse relaxation time and the mean equilibrium size of the chain in terms of the number of correlation blobs, it follows that
\[ {W\!i} = \lambda_{1} \dot \epsilon = \left( \lambda_\text{c} \, {\dot \epsilon} \right) \, \mathcal{N}_\text{c}^{2} = 
\left( \frac{\xi_\text{c} }{\xi_\text{P} } \right)^4  \,  \mathcal{N}_\text{c}^{2} = \left( \frac{R_\text{eq}}{\xi_\text{P} } \right)^4 \]
The size of the Pincus blob is consequently given by
\begin{equation}
\xi_\text{P}  = R_\text{eq} \, {W\!i}^{-\frac{1}{4}} 
\end{equation} 
Since the chain obeys RW statistics on all length scales,
\[ R_\text{eq} = \xi_\text{T} \mathcal{N}_\text{T}^{1/2} = \xi_\text{T}  \, z \]
As a result, the Pincus blob size becomes equal to the size of the thermal blob when
\[ W\!i_\text{c} = z^{4} \]
\vskip10pt
\noindent (b) \underline{$\xi_\text{c} < \xi_\text{P} < \xi_\text{T}$}
\vskip10pt
In this case as well, using arguments similar to those above, one can show that
\begin{equation}
\xi_\text{P}  = R_\text{eq} \, {W\!i}^{-\frac{1}{4}} 
\end{equation} 
In this concentration regime~\cite{JainPRL}, 
\[ \mathcal{N}_\text{c} =   \left( \frac{c}{c^*} \right)^{2} \]
As a result, since $R_\text{eq} = \xi_\text{c}  \mathcal{N}_\text{c}^{1/2} = 
\xi_\text{c} \, (c/c^*)$, it follows that the Pincus and correlation blobs become equal in size at a critical Weissenberg number given by
\[ W\!i_\text{c} = \left( \frac{c}{c^*} \right)^{4} \]
\vskip10pt
\noindent (c) \underline{$\xi_\text{P} < \xi_\text{c} < \xi_\text{T}$}
\vskip10pt
Since Zimm dynamics are obeyed below $\xi_\text{c}$, we have $\xi_\text{P} = b_\text{K} \, g_\text{P}^{{1}/{2}}$, and the Zimm relaxation time of the Pincus blob is $\lambda_\text{P} = \lambda_{0} \, g_\text{P}^{{3}/{2}}$. The flow penetrates a Pincus blob when $\lambda_\text{P} = {\dot \epsilon}^{-1}$. As a result
\[  \left( \frac{\xi_\text{P} }{b_\text{K}} \right)^3 = \frac{\lambda_\text{P} }{\lambda_{0}} = \left( \lambda_{0} \, {\dot \epsilon} \right)^{-1} \]
If there are $g_\text{c}$ monomers in a correlation blob, then $\xi_\text{c} = b_\text{K}\, g_\text{c}^{1/2}$, and the Rouse relaxation time of a correlation blob is $\lambda_\text{c} = \lambda_0 \, g_\text{c}^{3/2}$. As a result
\[  \left( \frac{\xi_\text{c} }{b_\text{K} } \right)^3 = \frac{\lambda_\text{c} }{\lambda_0}  \]
Clearly
\[  \frac{\lambda_1}{\lambda_\text{0}} = \frac{\lambda_1}{\lambda_\text{c}} \frac{\lambda_\text{c}}{\lambda_0} = \mathcal{N}_\text{c}^{2} \left( \frac{\xi_\text{c} }{b_\text{K}} \right)^3  \]
It follows that
\begin{align*}  
{W\!i} & = \lambda_{1} \dot \epsilon = (\lambda_\text{0} {\dot \epsilon})  \left( \frac{\lambda_1}{\lambda_\text{0}} \right) \nonumber \\ 
& =  \left( \frac{b_\text{K}}{\xi_\text{P} } \right)^3 \left( \frac{\xi_\text{c}}{b_\text{K}} \right)^3 \mathcal{N}_\text{c}^{2} 
= \left( \frac{\xi_\text{c}}{\xi_\text{P} } \right)^3  \mathcal{N}_\text{c}^{2}
\end{align*}  
Since $\xi_\text{c}^3 \mathcal{N}_\text{c}^{2} = R_\text{eq}^3 (c/c^*)$, the size of the Pincus blob depends on the Weissenberg number through
\[ \xi_\text{P}  = {W\!i}^{-\frac{1}{3}} \, R_\text{eq} \left( \frac{c}{c^*} \right)^{\frac{1}{3}} \]
RW statistics at all length scales implies $ R_\text{eq} = b_\text{K} N_\text{K}^{1/2}$. Consequently, at high Weissenberg numbers, when $\xi_\text{P} = b_\text{K}$, the successive fine graining procedure is expected to breakdown at 
\[ W\!i = N_\text{K}^{3/2} \left( \frac{c}{c^*} \right) \] 

\end{appendix}



%
%

\pagebreak

\onecolumngrid 

\clearpage 


\begin{center}
\textbf{\large Supplementary Material}
\end{center}


\twocolumngrid 

\setcounter{equation}{0}
\setcounter{figure}{0}
\setcounter{table}{0}
\setcounter{section}{0}
\setcounter{page}{1}
\makeatletter
\renewcommand{\theequation}{S\arabic{equation}}
\renewcommand{\thefigure}{S\arabic{figure}}
\renewcommand{\bibnumfmt}[1]{[S#1]}
\renewcommand{\citenumfont}[1]{S#1} 


\section{\label{sec:BD}Brownian dynamics simulations}
\vskip-10pt
The Euler integration algorithm for the nondimensional Ito stochastic differential equation governing the position vector $\Vector{r}^*_{\nu} (t^*)$ of bead $\nu$ at time $t^*$ is
\begin{align}
\label{ee1}
\Vector{r}^*_\nu (t^* + \Delta t^*)  & =  \Vector{r}^*_\nu (t^*) + \left[\bm{\Tensor{\kappa^*}}\cdot \Vector{r}^*_\nu (t^*)\right]\Delta t^* \nonumber \\ 
& + \frac{\Delta t^*}{4}\sum_{\mu = 1}^{N}\left[\bm{\Tensor{D}}_{\nu \mu }(t^*)\cdot \Vector{F}^*_{\mu}(t^*)\right] \nonumber \\ 
& + \frac{1}{\sqrt{2}}\sum_{\mu =1 }^{N} \left[\bm{\Tensor{B}}_{\nu \mu}(t^*)\cdot \Vector{\Delta}\bm{W}_\mu (t^*)\right]
\end{align}
In the above equation, $\bm{\Tensor{\kappa}^*}$ is a time dependent and homogeneous velocity gradient tensor which is equal to $(\bm{\nabla}\bm{v}^*)^{T}$ with $\bm{v^*}$ being the unperturbed solvent velocity. The nondimensional diffusion tensor $\bm{\Tensor{D}}_{\nu \mu }$ is a $3 \times 3$ square matrix for a fixed pair of particles $\mu$ and $\nu$, which is related to the dimensionless hydrodynamic interaction tensor, $\bm{\Tensor{\varOmega}}$, as follows:
\begin{equation}
\label{ee2}
\bm{\Tensor{D}}_{\nu \mu} = \delta_{\nu \mu} \, \bm{\Tensor{\delta}} + (1-\delta_{\nu \mu}) \,\bm{\Tensor{\varOmega}}(\Vector{r}^*_{\nu}-\Vector{r}^*_{\mu})
\end{equation}
where, $\bm{\Tensor{\delta}}$ and $\delta_{\mu \nu}$ represent a unit tensor and Kronecker delta, respectively. The hydrodynamic interaction tensor is represented by the Rotne-Prager-Yamakawa (RPY) tensor, which is discussed in greater detail below. The force, $\Vector{F}^*_\nu$, incorporates all the non-hydrodynamic forces acting on bead $\nu$ due to the presence of all other beads, for instance, in the present case, these are the spring forces and excluded volume interaction forces, i.e., $\Vector{F}^*_{\nu} = {\Vector{F}^*_{\nu}}^\text{sp}+{\Vector{F}^*_{\nu}}^\text{ev}$, discussed in greater detail below. The term $\bm{\Tensor{B}}_{\nu \mu}$ is a nondimensional tensor which is responsible for multiplicative noise, and is evaluated by decomposing the diffusion tensor as follows:
\begin{equation}
\label{ee3}
\bcal{B}\cdot\bcal{B}^{\textsc{t}} = \bcal{D}
\end{equation}       
where $\bcal{B}$ and $\bcal{D}$ are the block matrices consisting of $N \times N$ blocks each having dimensions $3 \times 3$, with the $(\nu, \mu)$th block of $\bcal{D}$ containing the components of the diffusion tensor $\bm{\Tensor{D}}_{\nu \mu}$, and the corresponding block of $\bcal{B}$ being equal to $\bm{\Tensor{B}}_{\nu \mu}$. The decomposition of the diffusion tensor has been achieved with the help of Fixman's polynomial approximation based on the Chebyshev technique which has been widely used  earlier for both single chain~\cite{Sfixman1986implicit,Sprabhakar2004multiplicative,SPrabhakarSFG} and multi-chain BD simulations~\cite{SJainPRE,SStoltz,SSaadatSemi}. The components of the white noise $\bm{\Delta}\Vector{W}_{\nu}$ are obtained from a real-valued Gaussian distribution with zero mean and variance $\Delta t^*$.
 
The entropic spring force, ${\Vector{F}^*_{\nu}}^\text{sp}$, on bead $\nu$ due to adjacent beads can be expressed as ${\Vector{F}^*_{\nu}}^\text{sp} = {\Vector{F}^*_{\nu}}^{c} - {\Vector{F}^*}_{\nu-1}^{c}$, where ${\Vector{F}^*_{\nu}}^{c}$ is the connector force between the beads $\nu$ and $\nu+1$, acting in the direction of the connector vector between two subsequent beads, $\Vector{Q}^*_{\nu} = \Vector{r}^*_{\nu+1} - \Vector{r}^*_{\nu}$. A wormlike chain model, widely used to represent a variety of molecules ranging from biomacromolecules like DNA to wormlike micelles~\cite{SBastamante}, is used to represent the spring force acting between two beads. In nondimensional form, it is written as~\cite{SMarko} 
\begin{equation}
\label{ee4}
{\Vector{F}^*}^{c}_\text{WLC}(\Vector{Q^*}) = \frac{1}{6q^*}\left( 4q^* + \frac{1}{(1-q^*)^{2}} - 1\right)\Vector{Q^*}
\end{equation}
In the above equation, $q^* = Q^*/\sqrt{b}$, where $Q^*$ is the magnitude of the nondimensional connector vector $\Vector{Q^*}$, and $b = Q_0^2/l_H^2$ is the finite extensibility parameter, with  $Q_0$ being the fully stretched length of the dimensional connector vector, $\Vector{Q}$.  

The excluded volume force ${\Vector{F^*}}^\text{ev}$ is given by
\begin{equation}
\label{ExvEq}
{\Vector{F^*}}^\text{ev} (\Vector{r}^{*}) = -\frac{\partial E (\Vector{r}^{*}) }{\partial \Vector{r}^{*}}
\end{equation}
where the narrow Gaussian potential discussed in section~II is used as the excluded volume potential $E (\Vector{r}^{*})$.

The hydrodynamic interaction tensor $\bm{\Tensor{\varOmega}}$ is given by the Rotne-Prager-Yamakawa (RPY) tensor~\cite{SRotnePrager,SYamakawa} which is a regularization of the Oseen-Burgers tensor written in nondimensional form as,
\begin{equation}
\label{ee5}
\bm{\Tensor{\varOmega}}(\Vector{r^*}) = \varOmega_{1} \, \bm{\Tensor{\delta}} + \varOmega_{2} \, \frac{\Vector{r^*}\, \Vector{r^*}}{{{r}^*}^2}
\end{equation}      
where $\Vector{r^*}$ is the separation distance vector between two beads, and ${r}^*$ is its magnitude.
For ${r}^*\geq 2a^{*}$, the functions $\varOmega_1$ and $\varOmega_2$ are given by
\begin{equation}
\label{ee6}
\varOmega_{1} = \frac{3a^{*}}{4r^*}\left(1+\frac{2{a^{*}}^{2}}{3{r^*}^{2}}\right) \, ; \quad  \Omega_{2} = \frac{3a^{*}}{4r^*}\left(1-\frac{2{a^{*}}^{2}}{{r^*}^{2}}\right)  
\end{equation} 
while for $0 < r^* \le 2 a^{*}$, they are given by
\begin{equation}
\label{ee7}
\varOmega_{1} = \left(1-\frac{9}{32}\frac{r^*}{a^{*}}\right)  \, ; \quad  \varOmega_{2} = \left(\frac{3}{32}\frac{r^*}{a^{*}}\right)  
\end{equation}
In the above expressions, $a^{*}$ is the nondimensional particle radius which is related to the conventionally defined hydrodynamic interaction parameter $h^*$ by $a^{*} = \sqrt{\pi} h^*$. It is well known that the sum, $\sum_{\mu}\bm{\Tensor{D}}_{\nu \mu }\cdot \Vector{F}^*_{\mu} $, in Eq.~(\ref{ee1}) is a conditionally convergent sum. Here it is evaluated using an optimized Ewald summation technique developed previously by Jain \etal~\cite{SJainPRE}.
 \begin{figure}[t]
\centering
\includegraphics[width=8.5cm,height=!]{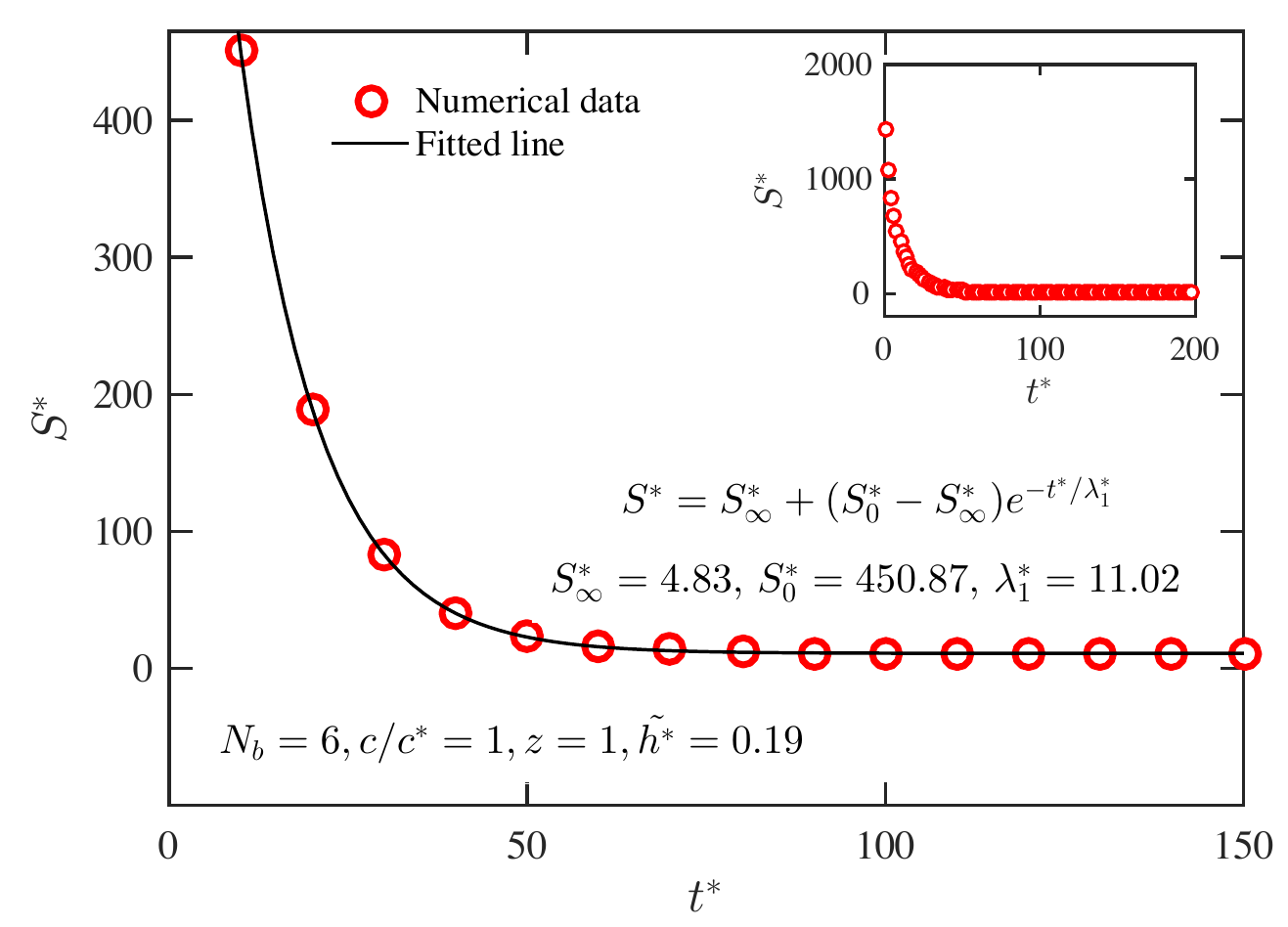}
\caption{\footnotesize Illustration of the procedure for finding the longest relaxation time $\lambda_1^*$ for $N_{b} = 6$. Error bars are smaller than symbol size. Inset shows the nondimensional mean-square stretch as a function of time over the course of the complete simulation. Parameter values for the simulation are:  $c/c^* = 1,\, z = 1, \,\tilde{h}^* = 0.19$, and $ N_{k} = 200$.}
\label{RelaxFit}
\end{figure}

The initial simulation box size $L_\text{sim}$ is selected such that $L_\text{sim} \ge 2R_e$, in order to prevent a chain from wrapping over itself. For this purpose, the value of $R_e$ at any value of $c/c^\star$ and $N_b$ is estimated from the blob scaling law $R_e = R_{e}^{0} \, (c/c^\star)^{(2\nu - 1)/(2-6\nu)}$, where $R_e^0$ is the end-to-end distance of a chain computed in the dilute limit. The BD algorithm used here adopts a number of modules from the ``The Molecular Modelling Toolkit'' (MMTK), which is an Open Source program library for molecular simulation applications~\cite{Smmtk}. For instance, initial configurations of the polymer chains were obtained by placing them randomly between $- L_\text{sim}/2$ and $+ L_\text{sim}/2$, with the help of the subroutine \textsf{randomPointInBox} from the MMTK package, which ensures that they do not overlap with each other. After calculating the initial simulation box size and generating the initial configurations of the chains, they are allowed to relax in the ``equilibration stage" for at least 100 non-dimensional time units with the flow turned off. By plotting static properties such as $\avg{R_e^2}$ and $\avg{R_g^2}$, which are expected to reach steady state after equilibration, we make sure that the system is at equilibrium. Following the equilibration run, we perform production runs (with the flow turned on), for varying lengths of time depending upon the value of strain up to which the experiments were carried out. Subsequent to a time convergence study, a non-dimensional time step size of 0.005 has been used in both the equilibration and production runs in all the cases considered here. Moreover, data were obtained by averaging over 50-70 independent runs, where each run contains around 60 chains in the simulation box at $c/c^* = 1$.

\begin{figure*}[t]
\centering
\includegraphics[width=1.0\textwidth]{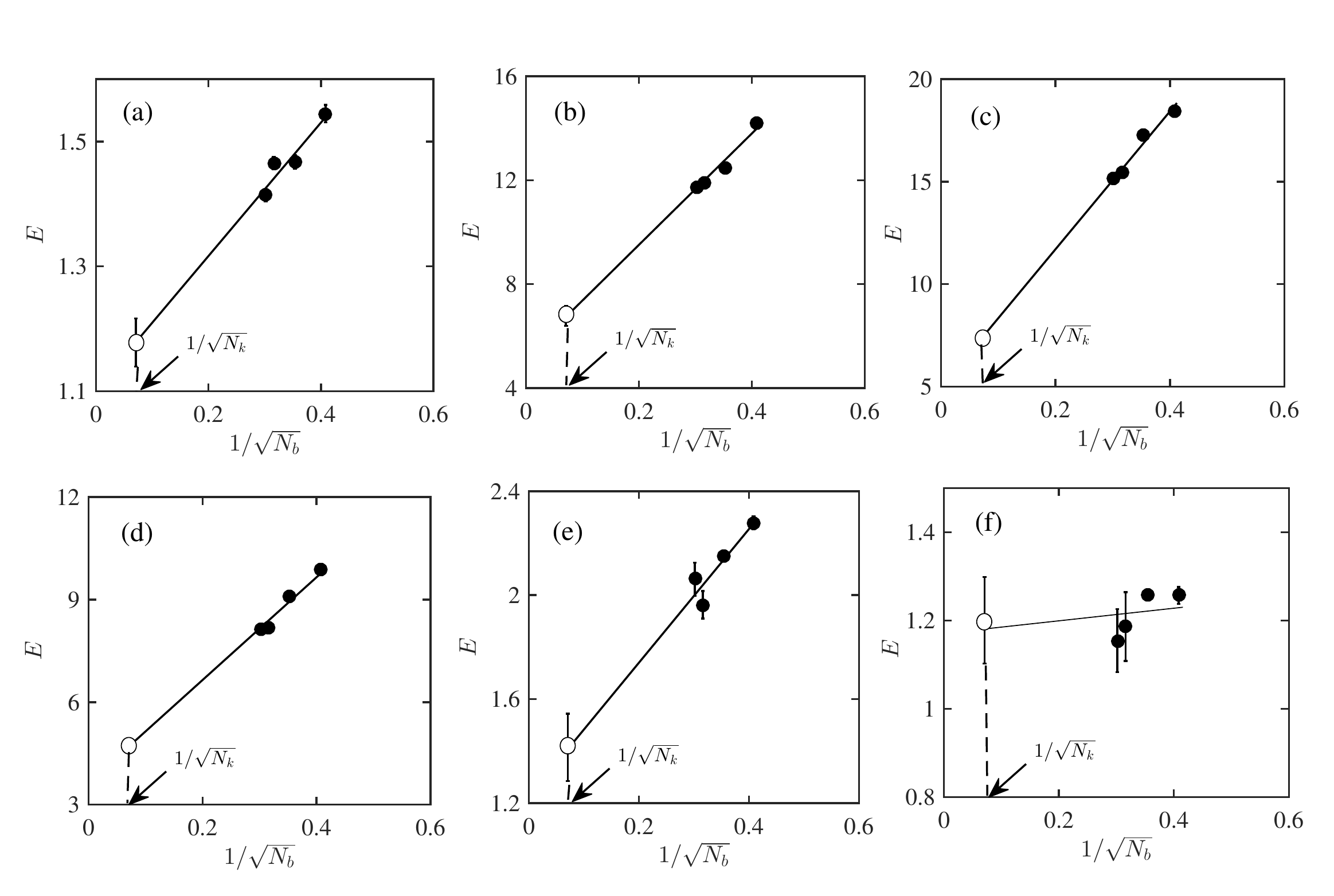}
\caption{ \footnotesize Results of the extrapolation procedure  at $z = 0.7$ during the stretching phase ((a) $\epsilon = 0.5$, (b) $\epsilon = 5$, and (c) $\epsilon = 11$), and the relaxation phase ((e) $t/\lambda_1 = 0.5$, $t/\lambda_1 = 2.5$, and $t/\lambda_1 = 4.5$). Filled symbols are the results of simulations while open symbols are the extrapolated values.  Parameters that are common to all simulations are: $c/c^* = 1,\, h^* = 0.19,$\, $ N_{k} = 200$ and $W\!i = 2.6$. Values of $b$, $\chi(b)$,  $z^*$, $\lambda_1^*$ and $\dot \epsilon^*$ used for each of the simulated values of $N_b = \{6, 8, 10, 12 \}$, are calculated as per the procedure described in section III of the main paper. Lines through the data at these values of $N_b$ indicate extrapolation to the limit $1/\sqrt{200}$.}
\label{Extraz0.7}
\centering
\includegraphics[width=0.9\textwidth]{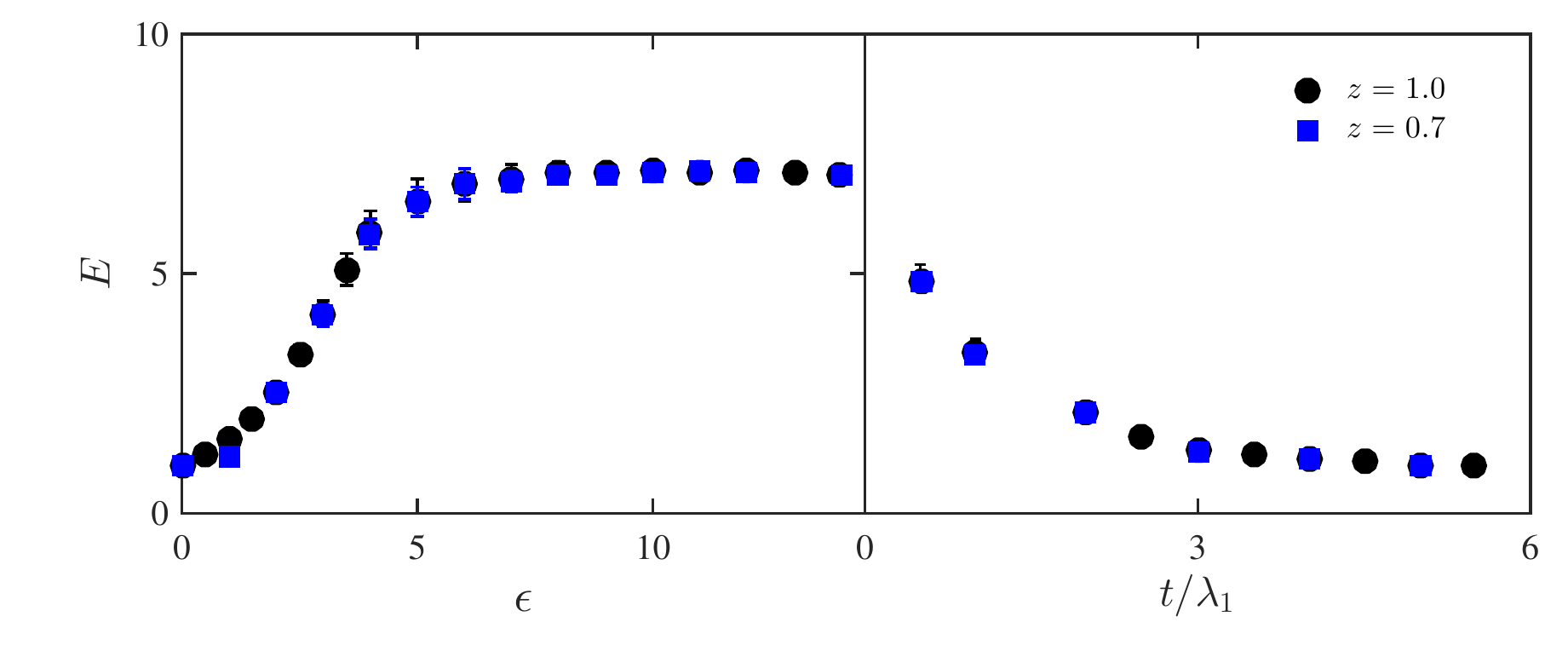}
\caption{ \footnotesize Comparison of the Expansion factor $E$ of a step strain followed by cessation of flow simulation for two values of $z$. Parameters that are common to all simulations are: $c/c^* = 1,$\, $ N_{k} = 200$, $\tilde{h^*} = 0.19$ and $W\!i = 2.6$.}
\label{FinalgraphdifferentZ}
\end{figure*}

A majority of the computer simulations were carried out on an IBM iDataplex x86 system (Barcoo) at the Victorian Life Sciences Computation Initiative (VLSCI). The system runs the RHEL 6 operating system, which is a variety of Linux, with compute nodes that currently perform at 20 teraFLOPS, with Xeon Phi cards running nominally at 1 teraFLOP each. For a single time step, with 60 chains each having 12 beads in an initial simulation box size of 15.3 units in length, calculations at $c/c^\star = 1.0$ typically took $3.217$ (s), of which the fraction of time devoted to  the decomposition of the diffusion tensor was $1.347$ (s).

\section{Results and discussions}

\subsection{Determination of the longest relaxation time}

\begin{figure*}[!t]
\centering
\includegraphics[width=0.9\textwidth]{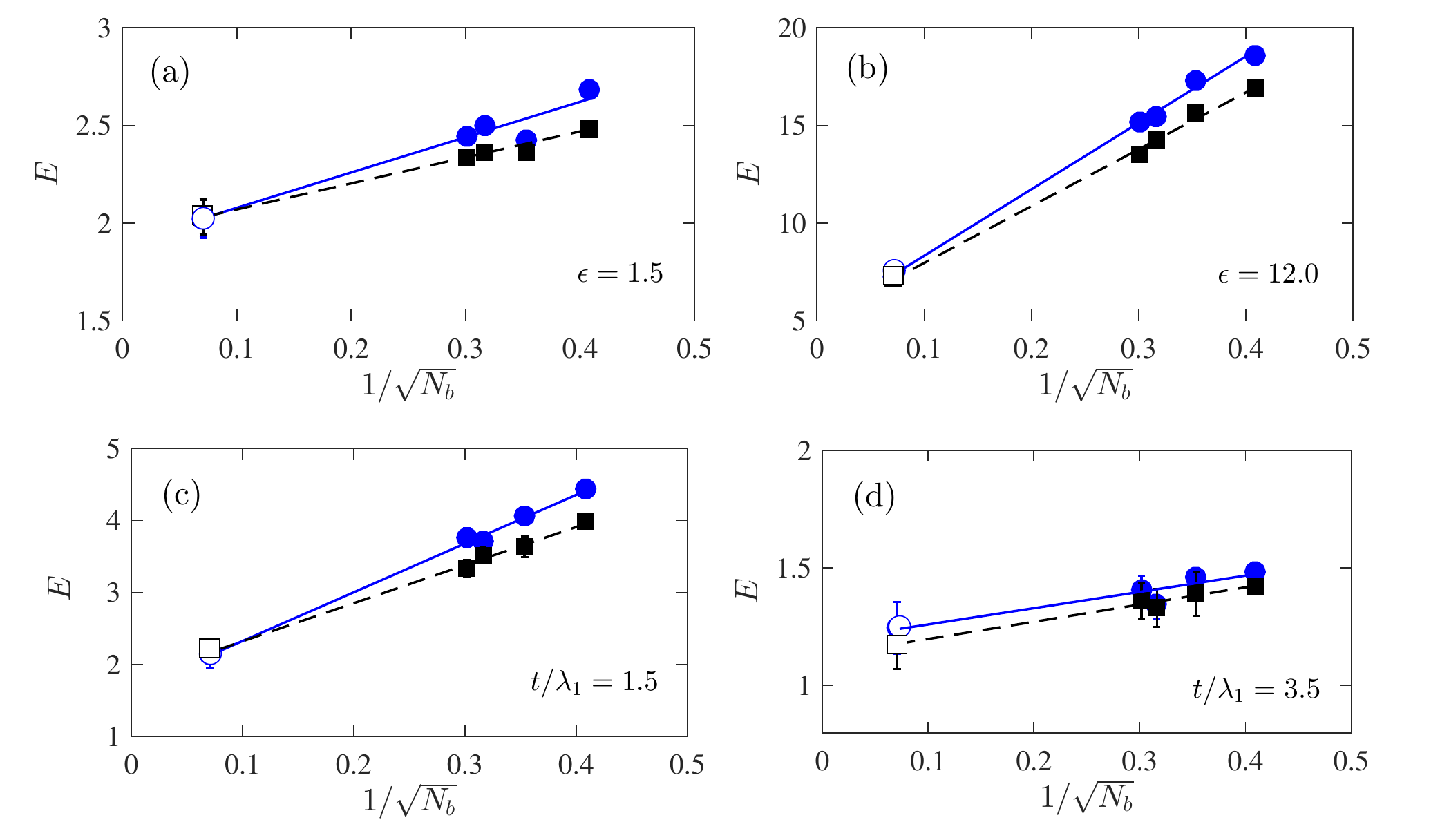}
\caption{ \footnotesize Results of the extrapolation procedure during the stretching phase ((a) $\epsilon = 1.5$, and~(b) $\epsilon = 12.0$), and the relaxation phase ((c) $t/\lambda_1 = 1.5$ and~(d) $t/\lambda_1 = 3.5$) of a step strain followed by cessation of flow simulation, for two values of $K$, namely, 1 (blue circle) and 1.5 (black square). Filled symbols are the results of simulations while open symbols are the extrapolated results.  Parameters that are common to all simulations are: $c/c^* = 1,\, z = 1,$\, $ N_{k} = 200$, $\tilde{h^*} = 0.19$ and $W\!i = 2.6$. Values of $b$, $\chi(b)$, $h^*$, $z^*$, $\lambda_1^*$ and $\dot \epsilon^*$ used for each of the simulated values of $N_b = \{6, 8, 10, 12 \}$, are calculated as per the procedure described in section III of the main paper. Lines through the data at these values of $N_b$ indicate extrapolation to the limit $1/\sqrt{200}$.}
\label{differentK}
\centering
\includegraphics[width=0.8\textwidth]{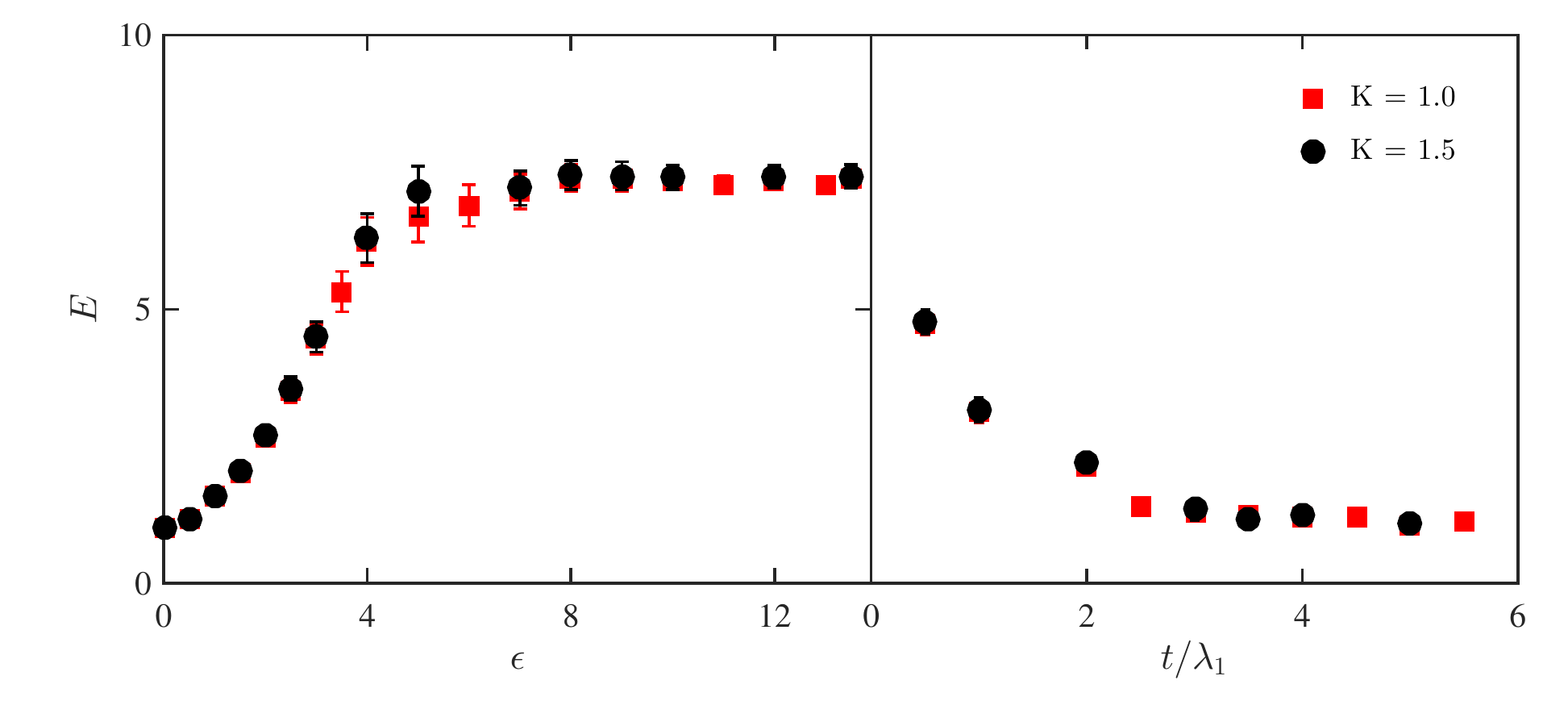}
\caption{ \footnotesize Comparison at two values of $K$ of the expansion factor $E$, for a step strain followed by cessation of flow simulation. Parameters that are common to all simulations are: $c/c^* = 1,\, z = 1,$\, $ N_{k} = 200$, $\tilde{h^*} = 0.19$ and $W\!i = 2.6$.}
\label{FinalgraphdifferentK}
\end{figure*} 

The longest dimensionless relaxation time $\lambda_1^*$, for any value of $N_b$, is obtained by initially stretching each chain to nearly 90$\%$ of its fully extended state, and letting it relax to equilibrium. The tail of the decay of the nondimensional mean-square stretch as a function of nondimensional time $t^*$ is then fitted with a single exponential function of the following form
\begin{equation}
\label{expfit}
S^* = {S}_\infty^* + \left( {S}_0^* -  {S}_\infty^* \right)\Exp{^{-t^* / \lambda_1^*}}
\end{equation}
where $S^*$ is the mean-square stretch $\avg{\bar{X}^*_\text{max} \cdot \bar{X}^*_\text{max}}$, and $ {S}_0^*$ and $ {S}_\infty^*$ are the initial and final (after the chain has fully relaxed) values of stretch, respectively, to which the fit is carried out. All three parameters, ${S}_0^*, {S}_\infty^*$ and $\lambda_1^*$ are determined from the fit. As expected, the value of $\sqrt{ {S}_\infty^*}$ is close to that of  $\bar{X}^*_\text{eq}$. However, it should be noted that the latter value is obtained from carrying out a static ensemble average from an equilibrium simulation, after the trajectories have reached a stationary state, as described in Ref.~\cite{SJainThesis}. 

An illustration of this procedure for $N_{b} = 6$ is given in Fig.~\ref{RelaxFit}, where the parameters  $c/c^* = 1,\, z = 1, \,\tilde{h}^* = 0.19$, and $ N_{k} = 200$ were used in the simulation. The inset shows the decay in the value of the dimensionless mean-square stretch as a function of time over the course of the complete simulation, while the main figure zooms in on the tail of the decay where the fit is carried out. The values of $\lambda_1^*$ obtained by this procedure for various values of $N_{b}$, which are required for carrying out the successive fine graining procedure, are given in Table~I of the main paper. 

\subsection{Comparison of stretch and relaxation predictions for $z = 0.7$ and $z = 1$}

The solvent quality has been chosen to be $z = 1$ in all our simulations, while as mentioned earlier, the measurements of Pan \etal\ \cite{Span2014universal,Span2014viscosity} suggest that a solution of $\lambda$-phage DNA has a solvent quality $z\approx 0.7$ at 22\degC. In order to test the assumption that the difference between results for these two values of $z$ will not be significant, simulations with $z=0.7$, have been carried out for the set of parameters: $W\!i = 2.6$, $c/c^* = 1.0$, $\tilde{h^*} = 0.19$ and $N_k = 200$. It should be noted that only one value of $\tilde{h^*}$ has been used for this purpose since the independence from choice of value for the hydrodynamic interaction parameter  has been established in Figs.~5 and~6 of the main paper. 

Results of the extrapolation procedure at $z = 0.7$, during both the stretch and relaxation phases at different values of $\epsilon$ and $t/\lambda_1$, are presented in Fig.~\ref{Extraz0.7} using the successive fine graining protocol outlined in the main paper. The final extrapolated results at both values of $z$, are compared in Fig.~\ref{FinalgraphdifferentZ}. It can be clearly seen that the predicted values are almost indistinguishable from each other at these two values of $z$. This nearly identical behaviour is expected to hold at all the values of $W\!i$ examined in this work. 

\subsection{Independence from the choice of value for $K$}

A truly parameter-free prediction of properties implies independence from the choice of the arbitrary constant $K$, since it only determines the range of the narrow Gaussian potential $d^*$, which is a local property. This has been demonstrated previously in the long chain limit, both at equilibrium and in shear flow for dilute solutions~\cite{Sprakash2001influence,SKumar,Skumar2004universal}, and at equilibrium for semidilute solutions~\cite{SJainPRL}. 

In order to establish independence from the range of the potential in the present case, which includes both flow and the cessation of flow, we carried out simulations for fixed values of $z = 1.0$, $c/c^* = 1.0$ and $Wi = 2.6$, and varied $d^\star$ according to $d^* = K{z^*}^{1/5}$, with two different values of $K$, namely, 1 and 1.5. The results are presented in Figs.~\ref{differentK} and~\ref{FinalgraphdifferentK}.  It is clear from Fig.~\ref{differentK} that the extrapolated results are virtually identical within error bars for these two values of $K$.  A comparison of the complete stretch and relaxation phases for the two values of $K$ is presented in Fig.~\ref{FinalgraphdifferentK}. Since the results are well in agreement for the two values of $K$, the simulation predictions can be considered to be free of the choice of the underlying parameters in the model.

\end{document}